%% file: main.tex
\PassOptionsToPackage{table}{xcolor}
\documentclass[conference,compsoc]{IEEEtran}

\input{sections/00-preamble}
\input{sections/00a-glossary}

\begin{document}

% Don't want date printed
\date{}

\title{``Please help share!'': Security and Privacy Advice on Twitter during the 2022 Russian Invasion of Ukraine }
\input{authors}

\maketitle

% Number files
\input{numbers/numbers-generated}
\input{numbers/numbers}

\begin{abstract}
\input{sections/01-abstract}
\end{abstract}

\section{Introduction}\label{sec:intro}
\input{sections/02-intro}

\section{Related Work}\label{sec:related-work}
\input{sections/04-related-work}

\section{Methodology}\label{sec:method}
\input{sections/05-method}

\section{Results}\label{sec:results}
\input{sections/06-results}

\section{Discussion}\label{sec:discussion}
\input{sections/07-discussion}

\section{Conclusion}\label{sec:conclusion}
\input{sections/08-conclusion}
\printbibliography
\appendices
\input{sections/09-appendix}

\end{document}

%% file: sections/00-preamble.tex
\usepackage[utf8]{inputenc} % should not be necessary with current LaTeX versions
\usepackage[T1]{fontenc}

\usepackage{xcolor}
\PassOptionsToPackage{table}{xcolor}
\usepackage{svg}
\usepackage{graphicx}
\usepackage{subcaption}
\usepackage{float}

\usepackage[]{hyperref}
\hypersetup{
    hidelinks,% To disable all colors + boxes
}

\usepackage{enumitem}

\usepackage{colortbl}
\usepackage{tabularx}
\usepackage{makecell}
\usepackage{booktabs}
\usepackage[para, flushleft]{threeparttable}

\newcolumntype{L}[1]{>{\raggedright\arraybackslash}p{#1}} % left fixed width
\newcolumntype{C}[1]{>{\centering\arraybackslash}p{#1}} % center fixed width
\newcolumntype{R}[1]{>{\raggedleft\arraybackslash}p{#1}} % flush right fixed width

\usepackage{graphics}
\usepackage{caption}
\usepackage{subcaption}
\usepackage{tikz}
\usetikzlibrary{shapes,arrows,fit,matrix,positioning,trees,decorations.pathreplacing,calc}
\usepackage[gen]{eurosym}

\usepackage[normalem]{ulem}
\usepackage{fontawesome}

\usepackage{enumitem}
\setlist[itemize]{leftmargin=0.5em,parsep=0em}

\newcommand{\ie}{i.\,e.}
\newcommand{\eg}{e.\,g.}% Include manually entered numbers here

\usepackage{xurl}
\PassOptionsToPackage{hyphens,spaces,obeyspaces}{url}

\usepackage[
    backend=biber,
    isbn=false,
    doi=false,
    url=false,
    eprint=false,
    mincitenames=1,
    maxcitenames=1,
    uniquelist=false,
    style=ieee]{biblatex}
\input{snippets/summary_boxes.tex}

\input{snippets/boldparagraphs.tex}

\input{snippets/quotes.tex}

\input{snippets/bettervariablenames.tex}

%% file: snippets/summary_boxes.tex
\usepackage[usetwoside=true]{mdframed}
\usepackage{tcolorbox}
\tcbuselibrary{breakable}
\tcbuselibrary{skins}

\definecolor{darkgray}{gray}{0.3}

\tcbset{}
\newtcolorbox{summaryBox}[2][]
{
    enhanced,
    breakable,
    frame hidden,
    fontupper       = \small,
    fontlower       = \footnotesize,
    borderline west = {2pt}{0pt}{lightgray},
    colback         = white,
    size            = fbox,
    left            = 0.5em,
    coltitle        = black,
    title           = {\color{darkgray}#2. },
    attach title to upper,
    #1,
}

%% file: snippets/boldparagraphs.tex
\makeatletter
\newcommand\boldparagraph{%
    \@startsection{paragraph}{5}%
    {\z@}{0.4ex plus 0.8ex minus 0.2ex}{0ex}%
    {\normalfont\normalsize\bfseries}%
}
\makeatother

%% file: snippets/quotes.tex
\usepackage[style=american,threshold=2]{csquotes}

%% file: snippets/bettervariablenames.tex
\usepackage{etoolbox}
\usepackage{xcolor}
\usepackage{ulem}

\newcommand\varunchanged[1]{{{\color{olive}#1}}}
\newcommand\varchanged[1]{{{\color{red}\sout{#1}}}}
\newcommand\varred[1]{{{\color{red}#1}}}

\newcommand\defineoldvar[2]{%
  \expandafter\newcommand\csname oldvar#1oldvar\endcsname{#2}%
}
\newcommand{\oldvar}[1]{\ifcsname oldvar#1oldvar\endcsname%
        \csname oldvar#1oldvar\endcsname%
    \else\PackageWarning{OldVar}{`#1' does not exist}{`#1' does not exist}%
        \varred{TODO}%
    \fi%
}

\newcommand\definevar[2]{%
    \expandafter\newcommand\csname var#1var\endcsname{#2}%
}
\newcommand{\var}[1]{\ifcsname var#1var\endcsname%
        \csname var#1var\endcsname%
    \else\PackageWarning{Var}{`#1' does not exist.}{`#1' does not exist.}%
        \varred{TODO}%
    \fi%
}

\newcommand{\compvar}[1]{\ifcsname var#1var\endcsname%
        \ifcsequal{var#1var}{oldvar#1oldvar}
        {\varunchanged{\csname var#1var\endcsname{} (U)}}
        {\csname var#1var\endcsname{} \varchanged{(\csname oldvar#1oldvar\endcsname)}}%
    \else\PackageWarning{VarComparison}{`#1' (NewVar) does not exist.}%
        \varred{TODO}%
    \fi%
    \ifcsname oldvar#1oldvar\endcsname%
    \else\PackageWarning{VarComparison}{`#1' (OldVar) does not exist.}%
    \fi%
}

%% file: sections/00a-glossary.tex
\usepackage[acronym]{glossaries}

\newglossaryentry{event}{%
    name={2022 Russian Invasion of Ukraine},
    description={} % Not displayed in paper
}
\newglossaryentry{event-short}{%
    name={invasion},
    description={} % Not displayed in paper
}
\newglossaryentry{tw}{%
    name={Twitter},
    description={} % Not displayed in paper
}

\newacronym{blm}{BLM}{Black Lives Matter}
\newacronym{dpr}{DPR}{Donetsk People's Republic}
\newacronym{lpr}{LPR}{Luhansk People's Republic}
\newacronym{mturk}{MTurk}{Amazon Mechanical Turk}
\newacronym{isp}{ISP}{Internet Service Provider}
\newacronym{ngo}{NGO}{non-government organization}
\newacronym{bgp}{BGP}{Border Gateway Protocol}
\newacronym{mh17}{MH17}{Malaysian Airlines Flight 17}
\newacronym{i2p}{I2P}{Invisible Internet Project}
\newacronym{2fa}{2FA}{two-factor authentication}
\newacronym{mfa}{MFA}{multi-factor authentication}
\newacronym{ddos}{DDoS}{distributed denial-of-service}
\newacronym{apt}{APT}{advanced persistent threat}
\newacronym{kw}{KW}{Kruskal-Wallis test}
\newacronym{mwu}{MWU}{Mann-Whitney U test}
\newacronym{irb}{IRB}{Institutional Review Board}
\newacronym{npo}{NPO}{nonprofit organization}

%% file: authors.tex
%% Authors
\author{
    \IEEEauthorblockN{%
        Juliane Schmüser\,\IEEEauthorrefmark{1},
        Noah Wöhler\,\IEEEauthorrefmark{1},
        Harshini Sri Ramulu\,\IEEEauthorrefmark{2},
        Christian Stransky\,\IEEEauthorrefmark{3},\\
        Dominik Wermke\,\IEEEauthorrefmark{1},
        Sascha Fahl\,\IEEEauthorrefmark{1}, and
        Yasemin Acar\,\IEEEauthorrefmark{2}
    }
    \IEEEauthorblockA{\IEEEauthorrefmark{1}CISPA Helmholtz Center for Information Security, Germany, \hypersetup{hidelinks}\\\texttt{\{\href{mailto:juliane.schmueser@cispa.de}{juliane.schmueser},\href{mailto:noah.woehler@cispa.de}{noah.woehler},\href{mailto:dominik.wermke@cispa.de}{dominik.wermke},\href{mailto:sascha.fahl@cispa.de}{sascha.fahl}\}@cispa.de}}
    \IEEEauthorblockA{\IEEEauthorrefmark{2}George Washington University, United States, \hypersetup{hidelinks}\texttt{\{\href{mailto:acar@gwu.edu}{acar}, \href{mailto:sharshini@gwu.edu}{sharshini}\}@gwu.edu}}
    \IEEEauthorblockA{\IEEEauthorrefmark{3}CISPA Helmholtz Center for Information Security, Germany, \hypersetup{hidelinks}\texttt{\href{mailto:stransky@sec.uni-hannover.de}stransky@sec.uni-hannover.de}}
}

%% file: numbers/numbers-generated.tex
% DO NOT include numbers manually here, this file will be overwritten
% (Add them to 'numbers.tex' instead)
% This file is generated by 'gen-numbers.py'
% VERSION: v0.3 	DATE: 2022-07-26

% Aggregates
\definevar{tweets.total}{8,920}
\definevar{tweets.relevant}{1,228}
\definevar{tweets.likes.median}{56}
\definevar{tweets.likes.sd}{5,151}
\definevar{tweets.retweets.median}{30}
\definevar{tweets.retweets.sd}{2,468}
\definevar{tweets.coded}{232}
\definevar{resources.total}{145}
\definevar{resources.relevant}{145}
\definevar{resources.coded}{140}
\definevar{tweets.total.text}{8,920}
\definevar{tweets.relevant.text}{1,228}
\definevar{tweets.likes.median.text}{56}
\definevar{tweets.likes.sd.text}{5,151}
\definevar{tweets.retweets.median.text}{30}
\definevar{tweets.retweets.sd.text}{2,468}
\definevar{tweets.coded.text}{244}
\definevar{resources.total.text}{145}
\definevar{resources.relevant.text}{145}
\definevar{resources.coded.text}{143}

% Codes
\definevar{codes.categories}{85}
\definevar{codes.unique}{507}
\definevar{codes.assigned.tweets}{888}
\definevar{codes.assigned.resources}{917}
\definevar{codes.unique.tweets}{161}
\definevar{codes.unique.resources}{212}
\definevar{codes.categories.text}{85}
\definevar{codes.unique.text}{507}
\definevar{codes.assigned.tweets.text}{888}
\definevar{codes.assigned.resources.text}{917}
\definevar{codes.unique.tweets.text}{161}
\definevar{codes.unique.resources.text}{212}

% CodeCountsBoth
\definevar{codeCountBoth.G1}{1}
\definevar{codeCountBoth.G1.1}{1}
\definevar{codeCountBoth.G2}{0}
\definevar{codeCountBoth.G2.1}{0}
\definevar{codeCountBoth.G2.2}{0}
\definevar{codeCountBoth.G3}{4}
\definevar{codeCountBoth.G3.1}{1}
\definevar{codeCountBoth.G3.2}{2}
\definevar{codeCountBoth.G3.3}{1}
\definevar{codeCountBoth.G3.4}{4}
\definevar{codeCountBoth.G3.5}{0}
\definevar{codeCountBoth.G4}{10}
\definevar{codeCountBoth.G4.1}{2}
\definevar{codeCountBoth.G4.2}{3}
\definevar{codeCountBoth.G4.3}{3}
\definevar{codeCountBoth.G4.4}{1}
\definevar{codeCountBoth.G5}{28}
\definevar{codeCountBoth.G5.1}{16}
\definevar{codeCountBoth.G5.2}{1}
\definevar{codeCountBoth.G5.3}{2}
\definevar{codeCountBoth.G5.4}{1}
\definevar{codeCountBoth.G5.5}{2}
\definevar{codeCountBoth.G6}{10}
\definevar{codeCountBoth.G6.1}{1}
\definevar{codeCountBoth.G7}{41}
\definevar{codeCountBoth.G7.1}{0}
\definevar{codeCountBoth.G8}{43}
\definevar{codeCountBoth.G9}{5}
\definevar{codeCountBoth.G10}{5}
\definevar{codeCountBoth.G11}{2}
\definevar{codeCountBoth.G12}{11}
\definevar{codeCountBoth.G12.1}{9}
\definevar{codeCountBoth.G12.2}{3}
\definevar{codeCountBoth.G12.3}{6}
\definevar{codeCountBoth.G12.4}{3}
\definevar{codeCountBoth.Ac1}{0}
\definevar{codeCountBoth.Ac1.1}{4}
\definevar{codeCountBoth.Ac1.2}{0}
\definevar{codeCountBoth.Ac1.3}{0}
\definevar{codeCountBoth.Ac1.4}{1}
\definevar{codeCountBoth.Ac1.5}{2}
\definevar{codeCountBoth.Ac1.6}{7}
\definevar{codeCountBoth.Ac1.7}{3}
\definevar{codeCountBoth.Au1}{0}
\definevar{codeCountBoth.Au1.1}{18}
\definevar{codeCountBoth.Au1.2}{34}
\definevar{codeCountBoth.Au1.3}{0}
\definevar{codeCountBoth.Au1.4}{11}
\definevar{codeCountBoth.Au1.5}{0}
\definevar{codeCountBoth.Au1.6}{2}
\definevar{codeCountBoth.Au1.7}{3}
\definevar{codeCountBoth.Au1.8}{1}
\definevar{codeCountBoth.Au1.9}{1}
\definevar{codeCountBoth.Au1.10}{2}
\definevar{codeCountBoth.Au2}{30}
\definevar{codeCountBoth.Au2.1}{1}
\definevar{codeCountBoth.Au2.2}{1}
\definevar{codeCountBoth.Au2.3}{20}
\definevar{codeCountBoth.Au3}{0}
\definevar{codeCountBoth.Au4}{0}
\definevar{codeCountBoth.Au4.1}{8}
\definevar{codeCountBoth.Au4.2}{7}
\definevar{codeCountBoth.Au4.3}{2}
\definevar{codeCountBoth.Au4.4}{2}
\definevar{codeCountBoth.L1}{0}
\definevar{codeCountBoth.L1.1}{0}
\definevar{codeCountBoth.L1.2}{5}
\definevar{codeCountBoth.L1.3}{4}
\definevar{codeCountBoth.L2}{2}
\definevar{codeCountBoth.L2.1}{4}
\definevar{codeCountBoth.L2.2}{0}
\definevar{codeCountBoth.L2.3}{0}
\definevar{codeCountBoth.L3}{1}
\definevar{codeCountBoth.L3.1}{2}
\definevar{codeCountBoth.L3.2}{3}
\definevar{codeCountBoth.L3.3}{9}
\definevar{codeCountBoth.I1}{0}
\definevar{codeCountBoth.I1.1}{7}
\definevar{codeCountBoth.I1.2}{0}
\definevar{codeCountBoth.I1.3}{0}
\definevar{codeCountBoth.I1.4}{0}
\definevar{codeCountBoth.I1.5}{1}
\definevar{codeCountBoth.I2}{0}
\definevar{codeCountBoth.I2.1}{2}
\definevar{codeCountBoth.I2.2}{12}
\definevar{codeCountBoth.I2.3}{0}
\definevar{codeCountBoth.I2.4}{0}
\definevar{codeCountBoth.I2.5}{2}
\definevar{codeCountBoth.I2.6}{4}
\definevar{codeCountBoth.I3}{0}
\definevar{codeCountBoth.I3.1}{1}
\definevar{codeCountBoth.I3.2}{1}
\definevar{codeCountBoth.I3.3}{17}
\definevar{codeCountBoth.I3.4}{1}
\definevar{codeCountBoth.I4}{0}
\definevar{codeCountBoth.I4.1}{0}
\definevar{codeCountBoth.I4.2}{0}
\definevar{codeCountBoth.I4.3}{0}
\definevar{codeCountBoth.I4.4}{2}
\definevar{codeCountBoth.I4.5}{1}
\definevar{codeCountBoth.I4.6}{0}
\definevar{codeCountBoth.I4.7}{1}
\definevar{codeCountBoth.I4.8}{8}
\definevar{codeCountBoth.I4.9}{3}
\definevar{codeCountBoth.I4.10}{1}
\definevar{codeCountBoth.I4.11}{0}
\definevar{codeCountBoth.I4.12}{1}
\definevar{codeCountBoth.I4.13}{1}
\definevar{codeCountBoth.I4.14}{7}
\definevar{codeCountBoth.I5}{0}
\definevar{codeCountBoth.I5.1}{1}
\definevar{codeCountBoth.I5.2}{0}
\definevar{codeCountBoth.I6}{0}
\definevar{codeCountBoth.I6.1}{0}
\definevar{codeCountBoth.I6.2}{0}
\definevar{codeCountBoth.I6.3}{0}
\definevar{codeCountBoth.I6.4}{0}
\definevar{codeCountBoth.I6.5}{0}
\definevar{codeCountBoth.I6.6}{0}
\definevar{codeCountBoth.I6.7}{0}
\definevar{codeCountBoth.I6.8}{0}
\definevar{codeCountBoth.I6.9}{0}
\definevar{codeCountBoth.I6.10}{0}
\definevar{codeCountBoth.I6.11}{0}
\definevar{codeCountBoth.I6.12}{0}
\definevar{codeCountBoth.I7}{0}
\definevar{codeCountBoth.I7.1}{0}
\definevar{codeCountBoth.I7.2}{0}
\definevar{codeCountBoth.I7.3}{0}
\definevar{codeCountBoth.I7.4}{0}
\definevar{codeCountBoth.I7.5}{0}
\definevar{codeCountBoth.I7.6}{0}
\definevar{codeCountBoth.I7.7}{2}
\definevar{codeCountBoth.I7.8}{1}
\definevar{codeCountBoth.I8}{0}
\definevar{codeCountBoth.I8.1}{0}
\definevar{codeCountBoth.I8.2}{0}
\definevar{codeCountBoth.I8.3}{0}
\definevar{codeCountBoth.I9}{0}
\definevar{codeCountBoth.I9.1}{1}
\definevar{codeCountBoth.I9.2}{0}
\definevar{codeCountBoth.I9.3}{0}
\definevar{codeCountBoth.I9.4}{0}
\definevar{codeCountBoth.I9.5}{0}
\definevar{codeCountBoth.I9.6}{0}
\definevar{codeCountBoth.I9.7}{0}
\definevar{codeCountBoth.I9.8}{1}
\definevar{codeCountBoth.I9.9}{0}
\definevar{codeCountBoth.I9.10}{0}
\definevar{codeCountBoth.I10}{0}
\definevar{codeCountBoth.I10.1}{0}
\definevar{codeCountBoth.I10.2}{0}
\definevar{codeCountBoth.I10.3}{0}
\definevar{codeCountBoth.I10.4}{0}
\definevar{codeCountBoth.I10.5}{1}
\definevar{codeCountBoth.I10.6}{1}
\definevar{codeCountBoth.I10.7}{0}
\definevar{codeCountBoth.I10.8}{0}
\definevar{codeCountBoth.I10.9}{0}
\definevar{codeCountBoth.I10.10}{0}
\definevar{codeCountBoth.M1}{0}
\definevar{codeCountBoth.M1.1}{10}
\definevar{codeCountBoth.M1.2}{5}
\definevar{codeCountBoth.M1.3}{5}
\definevar{codeCountBoth.M1.4}{7}
\definevar{codeCountBoth.M2}{0}
\definevar{codeCountBoth.M2.1}{6}
\definevar{codeCountBoth.M2.2}{2}
\definevar{codeCountBoth.M3}{0}
\definevar{codeCountBoth.M3.1}{5}
\definevar{codeCountBoth.M4}{0}
\definevar{codeCountBoth.M4.1}{1}
\definevar{codeCountBoth.M4.2}{1}
\definevar{codeCountBoth.M4.3}{15}
\definevar{codeCountBoth.M4.4}{1}
\definevar{codeCountBoth.M4.5}{9}
\definevar{codeCountBoth.M4.6}{4}
\definevar{codeCountBoth.M4.7}{1}
\definevar{codeCountBoth.M4.8}{1}
\definevar{codeCountBoth.M4.9}{6}
\definevar{codeCountBoth.M4.10}{4}
\definevar{codeCountBoth.M4.11}{2}
\definevar{codeCountBoth.M4.12}{1}
\definevar{codeCountBoth.M4.13}{7}
\definevar{codeCountBoth.M4.14}{2}
\definevar{codeCountBoth.M4.15}{2}
\definevar{codeCountBoth.M4.16}{1}
\definevar{codeCountBoth.M4.17}{1}
\definevar{codeCountBoth.E1}{0}
\definevar{codeCountBoth.E1.1}{1}
\definevar{codeCountBoth.E1.2}{1}
\definevar{codeCountBoth.E1.3}{2}
\definevar{codeCountBoth.E2}{0}
\definevar{codeCountBoth.E2.1}{3}
\definevar{codeCountBoth.E2.2}{2}
\definevar{codeCountBoth.E2.3}{1}
\definevar{codeCountBoth.E2.4}{1}
\definevar{codeCountBoth.E2.5}{19}
\definevar{codeCountBoth.E2.6}{0}
\definevar{codeCountBoth.E2.7}{1}
\definevar{codeCountBoth.E2.8}{0}
\definevar{codeCountBoth.E2.9}{0}
\definevar{codeCountBoth.E2.10}{6}
\definevar{codeCountBoth.E2.11}{2}
\definevar{codeCountBoth.E2.12}{1}
\definevar{codeCountBoth.E3}{0}
\definevar{codeCountBoth.E3.1}{3}
\definevar{codeCountBoth.E3.2}{0}
\definevar{codeCountBoth.E3.3}{10}
\definevar{codeCountBoth.E3.4}{0}
\definevar{codeCountBoth.E4}{0}
\definevar{codeCountBoth.E4.1}{11}
\definevar{codeCountBoth.E4.2}{0}
\definevar{codeCountBoth.E4.3}{5}
\definevar{codeCountBoth.E4.4}{0}
\definevar{codeCountBoth.E4.5}{1}
\definevar{codeCountBoth.E4.6}{0}
\definevar{codeCountBoth.E5}{0}
\definevar{codeCountBoth.E5.1}{0}
\definevar{codeCountBoth.E5.2}{0}
\definevar{codeCountBoth.E5.3}{0}
\definevar{codeCountBoth.E5.4}{0}
\definevar{codeCountBoth.E5.5}{0}
\definevar{codeCountBoth.E5.6}{1}
\definevar{codeCountBoth.E6}{0}
\definevar{codeCountBoth.E6.1}{0}
\definevar{codeCountBoth.E6.2}{0}
\definevar{codeCountBoth.E6.3}{0}
\definevar{codeCountBoth.E6.4}{0}
\definevar{codeCountBoth.E6.5}{0}
\definevar{codeCountBoth.E6.6}{0}
\definevar{codeCountBoth.E6.7}{0}
\definevar{codeCountBoth.E6.8}{1}
\definevar{codeCountBoth.E7}{0}
\definevar{codeCountBoth.E7.1}{0}
\definevar{codeCountBoth.E7.2}{0}
\definevar{codeCountBoth.E7.3}{0}
\definevar{codeCountBoth.E7.4}{0}
\definevar{codeCountBoth.E7.5}{0}
\definevar{codeCountBoth.Soc1}{0}
\definevar{codeCountBoth.Soc1.1}{5}
\definevar{codeCountBoth.Soc1.2}{6}
\definevar{codeCountBoth.Soc1.3}{0}
\definevar{codeCountBoth.Soc1.4}{1}
\definevar{codeCountBoth.Soc1.5}{7}
\definevar{codeCountBoth.Soc1.6}{0}
\definevar{codeCountBoth.Soc1.7}{4}
\definevar{codeCountBoth.Soc1.8}{4}
\definevar{codeCountBoth.Soc1.9}{1}
\definevar{codeCountBoth.Soc1.10}{2}
\definevar{codeCountBoth.Soc1.11}{4}
\definevar{codeCountBoth.Soc1.12}{1}
\definevar{codeCountBoth.Soc1.13}{0}
\definevar{codeCountBoth.Soc2}{0}
\definevar{codeCountBoth.Soc2.1}{0}
\definevar{codeCountBoth.Soc2.2}{0}
\definevar{codeCountBoth.Soc2.3}{0}
\definevar{codeCountBoth.Soc2.4}{0}
\definevar{codeCountBoth.Soc2.5}{0}
\definevar{codeCountBoth.Soc2.6}{3}
\definevar{codeCountBoth.Soc3}{0}
\definevar{codeCountBoth.Soc3.1}{12}
\definevar{codeCountBoth.Soc3.2}{1}
\definevar{codeCountBoth.Soc3.3}{7}
\definevar{codeCountBoth.Soc3.4}{4}
\definevar{codeCountBoth.Soc3.5}{4}
\definevar{codeCountBoth.Soc4}{0}
\definevar{codeCountBoth.Soc4.1}{1}
\definevar{codeCountBoth.C1}{0}
\definevar{codeCountBoth.C1.1}{0}
\definevar{codeCountBoth.C1.2}{0}
\definevar{codeCountBoth.C1.3}{0}
\definevar{codeCountBoth.C2}{0}
\definevar{codeCountBoth.C2.1}{0}
\definevar{codeCountBoth.C2.2}{0}
\definevar{codeCountBoth.C3}{0}
\definevar{codeCountBoth.C3.1}{0}
\definevar{codeCountBoth.C3.2}{0}
\definevar{codeCountBoth.A1}{0}
\definevar{codeCountBoth.A1.1}{10}
\definevar{codeCountBoth.A1.2}{5}
\definevar{codeCountBoth.A1.3}{0}
\definevar{codeCountBoth.A1.4}{0}
\definevar{codeCountBoth.A1.5}{0}
\definevar{codeCountBoth.A1.6}{0}
\definevar{codeCountBoth.A1.7}{0}
\definevar{codeCountBoth.A1.8}{1}
\definevar{codeCountBoth.A1.9}{2}
\definevar{codeCountBoth.A1.10}{1}
\definevar{codeCountBoth.A1.11}{3}
\definevar{codeCountBoth.A1.12}{1}
\definevar{codeCountBoth.A1.13}{1}
\definevar{codeCountBoth.A1.14}{4}
\definevar{codeCountBoth.A2}{0}
\definevar{codeCountBoth.A2.1}{11}
\definevar{codeCountBoth.A2.2}{3}
\definevar{codeCountBoth.A2.3}{0}
\definevar{codeCountBoth.A2.4}{0}
\definevar{codeCountBoth.A2.5}{0}
\definevar{codeCountBoth.A2.6}{1}
\definevar{codeCountBoth.A2.7}{4}
\definevar{codeCountBoth.A2.8}{2}
\definevar{codeCountBoth.A2.9}{0}
\definevar{codeCountBoth.A3}{0}
\definevar{codeCountBoth.A3.1}{24}
\definevar{codeCountBoth.A3.2}{0}
\definevar{codeCountBoth.A3.3}{0}
\definevar{codeCountBoth.A3.4}{0}
\definevar{codeCountBoth.A3.5}{0}
\definevar{codeCountBoth.Sec1}{0}
\definevar{codeCountBoth.Sec1.1}{8}
\definevar{codeCountBoth.Sec1.2}{0}
\definevar{codeCountBoth.Sec1.3}{0}
\definevar{codeCountBoth.Sec1.4}{1}
\definevar{codeCountBoth.Sec1.5}{11}
\definevar{codeCountBoth.Sec1.6}{1}
\definevar{codeCountBoth.Sec1.7}{1}
\definevar{codeCountBoth.Sec1.8}{3}
\definevar{codeCountBoth.Sec1.9}{1}
\definevar{codeCountBoth.Sec1.10}{1}
\definevar{codeCountBoth.Sec1.11}{2}
\definevar{codeCountBoth.Sec1.12}{3}
\definevar{codeCountBoth.Sec1.13}{2}
\definevar{codeCountBoth.Sec1.14}{2}
\definevar{codeCountBoth.Sec2}{0}
\definevar{codeCountBoth.Sec2.1}{34}
\definevar{codeCountBoth.Sec2.2}{5}
\definevar{codeCountBoth.Sec2.3}{1}
\definevar{codeCountBoth.Sec2.4}{3}
\definevar{codeCountBoth.Sec2.5}{1}
\definevar{codeCountBoth.DS1}{0}
\definevar{codeCountBoth.DS1.1}{2}
\definevar{codeCountBoth.DS1.2}{1}
\definevar{codeCountBoth.DS2}{0}
\definevar{codeCountBoth.DS2.1}{0}
\definevar{codeCountBoth.DS2.2}{0}
\definevar{codeCountBoth.DS2.3}{1}
\definevar{codeCountBoth.DS2.4}{0}
\definevar{codeCountBoth.DS2.5}{0}
\definevar{codeCountBoth.DS2.6}{0}
\definevar{codeCountBoth.DS2.7}{0}
\definevar{codeCountBoth.DS2.8}{0}
\definevar{codeCountBoth.DS2.9}{2}
\definevar{codeCountBoth.DS2.10}{2}
\definevar{codeCountBoth.DS2.11}{1}
\definevar{codeCountBoth.DS3}{0}
\definevar{codeCountBoth.DS3.1}{0}
\definevar{codeCountBoth.DS3.2}{0}
\definevar{codeCountBoth.DS3.3}{3}
\definevar{codeCountBoth.DS3.4}{1}
\definevar{codeCountBoth.DS3.5}{0}
\definevar{codeCountBoth.DS3.6}{2}
\definevar{codeCountBoth.DS3.7}{1}
\definevar{codeCountBoth.DS3.8}{0}
\definevar{codeCountBoth.DS3.9}{0}
\definevar{codeCountBoth.DS3.10}{0}
\definevar{codeCountBoth.DS3.11}{0}
\definevar{codeCountBoth.DS3.12}{1}
\definevar{codeCountBoth.DS3.13}{0}
\definevar{codeCountBoth.DS3.14}{0}
\definevar{codeCountBoth.DS3.15}{3}
\definevar{codeCountBoth.DS3.16}{3}
\definevar{codeCountBoth.DS3.17}{0}
\definevar{codeCountBoth.DS3.18}{11}
\definevar{codeCountBoth.DS3.19}{1}
\definevar{codeCountBoth.DS3.20}{1}
\definevar{codeCountBoth.DS3.21}{2}
\definevar{codeCountBoth.DS3.22}{3}
\definevar{codeCountBoth.DS3.23}{3}
\definevar{codeCountBoth.DS3.24}{1}
\definevar{codeCountBoth.DS3.25}{1}
\definevar{codeCountBoth.DS3.26}{1}
\definevar{codeCountBoth.DS4}{0}
\definevar{codeCountBoth.DS4.1}{0}
\definevar{codeCountBoth.DS4.2}{0}
\definevar{codeCountBoth.DS4.3}{0}
\definevar{codeCountBoth.DS5}{0}
\definevar{codeCountBoth.DS5.1}{1}
\definevar{codeCountBoth.DS5.2}{0}
\definevar{codeCountBoth.DS5.3}{0}
\definevar{codeCountBoth.DS5.4}{0}
\definevar{codeCountBoth.DS5.5}{2}
\definevar{codeCountBoth.DS5.6}{10}
\definevar{codeCountBoth.DS5.7}{0}
\definevar{codeCountBoth.DS5.8}{1}
\definevar{codeCountBoth.DS5.9}{1}
\definevar{codeCountBoth.DS5.10}{1}
\definevar{codeCountBoth.DS5.11}{1}
\definevar{codeCountBoth.DS5.12}{2}
\definevar{codeCountBoth.DS5.13}{1}
\definevar{codeCountBoth.DS5.14}{1}
\definevar{codeCountBoth.DS5.15}{6}
\definevar{codeCountBoth.DS5.16}{4}
\definevar{codeCountBoth.DS5.17}{0}
\definevar{codeCountBoth.DS5.18}{1}
\definevar{codeCountBoth.DS5.19}{2}
\definevar{codeCountBoth.DS6}{0}
\definevar{codeCountBoth.DS6.1}{0}
\definevar{codeCountBoth.DS6.2}{0}
\definevar{codeCountBoth.DS6.3}{0}
\definevar{codeCountBoth.Sto1}{0}
\definevar{codeCountBoth.Sto1.1}{1}
\definevar{codeCountBoth.Sto1.2}{22}
\definevar{codeCountBoth.Sto1.3}{4}
\definevar{codeCountBoth.Sto1.4}{13}
\definevar{codeCountBoth.Sto1.5}{2}
\definevar{codeCountBoth.Sto1.6}{1}
\definevar{codeCountBoth.Sto1.7}{2}
\definevar{codeCountBoth.Sto1.8}{1}
\definevar{codeCountBoth.Sto1.9}{6}
\definevar{codeCountBoth.Sto1.10}{1}
\definevar{codeCountBoth.Sto2}{0}
\definevar{codeCountBoth.Sto2.1}{1}
\definevar{codeCountBoth.Sto2.2}{0}
\definevar{codeCountBoth.Sto2.3}{0}
\definevar{codeCountBoth.Sto2.4}{0}
\definevar{codeCountBoth.Sto2.5}{0}
\definevar{codeCountBoth.Sto2.6}{1}
\definevar{codeCountBoth.Sto2.7}{1}
\definevar{codeCountBoth.Sto3}{0}
\definevar{codeCountBoth.Sto3.1}{2}
\definevar{codeCountBoth.Sto3.2}{0}
\definevar{codeCountBoth.Sto3.3}{4}
\definevar{codeCountBoth.Sto3.4}{5}
\definevar{codeCountBoth.Sto3.5}{2}
\definevar{codeCountBoth.Sto3.6}{0}
\definevar{codeCountBoth.Sto3.7}{0}
\definevar{codeCountBoth.Sto3.8}{0}
\definevar{codeCountBoth.Sto3.9}{1}
\definevar{codeCountBoth.Sto4}{0}
\definevar{codeCountBoth.Sto4.1}{0}
\definevar{codeCountBoth.Sto4.2}{4}
\definevar{codeCountBoth.Sto4.3}{0}
\definevar{codeCountBoth.Sto4.4}{0}
\definevar{codeCountBoth.Sto5}{0}
\definevar{codeCountBoth.Sto5.1}{2}
\definevar{codeCountBoth.Sto5.2}{4}
\definevar{codeCountBoth.Sto5.3}{1}
\definevar{codeCountBoth.Sto5.4}{4}
\definevar{codeCountBoth.Sto5.5}{1}
\definevar{codeCountBoth.Sto5.6}{7}
\definevar{codeCountBoth.B1}{0}
\definevar{codeCountBoth.B1.1}{0}
\definevar{codeCountBoth.B1.2}{0}
\definevar{codeCountBoth.B1.3}{0}
\definevar{codeCountBoth.B2}{0}
\definevar{codeCountBoth.B2.1}{0}
\definevar{codeCountBoth.B2.2}{0}
\definevar{codeCountBoth.B2.3}{0}
\definevar{codeCountBoth.B2.4}{0}
\definevar{codeCountBoth.B2.5}{0}
\definevar{codeCountBoth.B2.6}{1}
\definevar{codeCountBoth.B2.7}{0}
\definevar{codeCountBoth.B3}{0}
\definevar{codeCountBoth.B3.1}{1}
\definevar{codeCountBoth.B3.2}{0}
\definevar{codeCountBoth.B3.3}{2}
\definevar{codeCountBoth.B3.4}{0}
\definevar{codeCountBoth.B3.5}{0}
\definevar{codeCountBoth.B3.6}{0}
\definevar{codeCountBoth.B3.7}{0}
\definevar{codeCountBoth.T1}{221}
\definevar{codeCountBoth.T2}{135}
\definevar{codeCountBoth.Co1}{0}
\definevar{codeCountBoth.Co1.1}{2}
\definevar{codeCountBoth.Re1}{1}
\definevar{codeCountBoth.Re1.1}{2}
\definevar{codeCountBoth.Re1.2}{6}
\definevar{codeCountBoth.Re1.3}{1}
\definevar{codeCountBoth.Re1.4}{1}
\definevar{codeCountBoth.S}{0}
\definevar{codeCountBoth.S1}{51}
\definevar{codeCountBoth.S1.1}{0}
\definevar{codeCountBoth.S1.2}{0}
\definevar{codeCountBoth.S1.3}{6}
\definevar{codeCountBoth.S1.4}{2}
\definevar{codeCountBoth.S1.5}{0}
\definevar{codeCountBoth.S2}{37}
\definevar{codeCountBoth.S2.1}{2}
\definevar{codeCountBoth.S2.2}{0}
\definevar{codeCountBoth.S2.3}{8}
\definevar{codeCountBoth.S2.4}{3}
\definevar{codeCountBoth.S2.5}{1}
\definevar{codeCountBoth.S3}{0}
\definevar{codeCountBoth.S3.1}{0}
\definevar{codeCountBoth.S3.2}{0}
\definevar{codeCountBoth.S3.3}{0}
\definevar{codeCountBoth.S3.4}{0}
\definevar{codeCountBoth.S3.5}{0}
\definevar{codeCountBoth.S4}{56}
\definevar{codeCountBoth.S4.1}{3}
\definevar{codeCountBoth.S4.2}{0}
\definevar{codeCountBoth.S4.3}{0}
\definevar{codeCountBoth.S4.4}{0}
\definevar{codeCountBoth.S4.5}{0}
\definevar{codeCountBoth.S5}{35}
\definevar{codeCountBoth.S5.1}{14}
\definevar{codeCountBoth.S5.2}{0}
\definevar{codeCountBoth.S5.3}{3}
\definevar{codeCountBoth.S5.4}{13}
\definevar{codeCountBoth.S5.5}{5}
\definevar{codeCountBoth.S6}{69}
\definevar{codeCountBoth.S6.1}{18}
\definevar{codeCountBoth.S6.2}{5}
\definevar{codeCountBoth.S6.3}{1}
\definevar{codeCountBoth.S6.4}{20}
\definevar{codeCountBoth.S6.5}{22}
\definevar{codeCountBoth.G1.text}{one}
\definevar{codeCountBoth.G1.1.text}{one}
\definevar{codeCountBoth.G2.text}{zero}
\definevar{codeCountBoth.G2.1.text}{zero}
\definevar{codeCountBoth.G2.2.text}{zero}
\definevar{codeCountBoth.G3.text}{four}
\definevar{codeCountBoth.G3.1.text}{one}
\definevar{codeCountBoth.G3.2.text}{two}
\definevar{codeCountBoth.G3.3.text}{one}
\definevar{codeCountBoth.G3.4.text}{four}
\definevar{codeCountBoth.G3.5.text}{zero}
\definevar{codeCountBoth.G4.text}{10}
\definevar{codeCountBoth.G4.1.text}{two}
\definevar{codeCountBoth.G4.2.text}{three}
\definevar{codeCountBoth.G4.3.text}{three}
\definevar{codeCountBoth.G4.4.text}{one}
\definevar{codeCountBoth.G5.text}{28}
\definevar{codeCountBoth.G5.1.text}{16}
\definevar{codeCountBoth.G5.2.text}{one}
\definevar{codeCountBoth.G5.3.text}{two}
\definevar{codeCountBoth.G5.4.text}{one}
\definevar{codeCountBoth.G5.5.text}{two}
\definevar{codeCountBoth.G6.text}{10}
\definevar{codeCountBoth.G6.1.text}{one}
\definevar{codeCountBoth.G7.text}{41}
\definevar{codeCountBoth.G7.1.text}{zero}
\definevar{codeCountBoth.G8.text}{43}
\definevar{codeCountBoth.G9.text}{five}
\definevar{codeCountBoth.G10.text}{five}
\definevar{codeCountBoth.G11.text}{two}
\definevar{codeCountBoth.G12.text}{11}
\definevar{codeCountBoth.G12.1.text}{nine}
\definevar{codeCountBoth.G12.2.text}{three}
\definevar{codeCountBoth.G12.3.text}{six}
\definevar{codeCountBoth.G12.4.text}{three}
\definevar{codeCountBoth.Ac1.text}{zero}
\definevar{codeCountBoth.Ac1.1.text}{four}
\definevar{codeCountBoth.Ac1.2.text}{zero}
\definevar{codeCountBoth.Ac1.3.text}{zero}
\definevar{codeCountBoth.Ac1.4.text}{one}
\definevar{codeCountBoth.Ac1.5.text}{two}
\definevar{codeCountBoth.Ac1.6.text}{seven}
\definevar{codeCountBoth.Ac1.7.text}{three}
\definevar{codeCountBoth.Au1.text}{zero}
\definevar{codeCountBoth.Au1.1.text}{18}
\definevar{codeCountBoth.Au1.2.text}{34}
\definevar{codeCountBoth.Au1.3.text}{zero}
\definevar{codeCountBoth.Au1.4.text}{11}
\definevar{codeCountBoth.Au1.5.text}{zero}
\definevar{codeCountBoth.Au1.6.text}{two}
\definevar{codeCountBoth.Au1.7.text}{three}
\definevar{codeCountBoth.Au1.8.text}{one}
\definevar{codeCountBoth.Au1.9.text}{one}
\definevar{codeCountBoth.Au1.10.text}{two}
\definevar{codeCountBoth.Au2.text}{30}
\definevar{codeCountBoth.Au2.1.text}{one}
\definevar{codeCountBoth.Au2.2.text}{one}
\definevar{codeCountBoth.Au2.3.text}{20}
\definevar{codeCountBoth.Au3.text}{zero}
\definevar{codeCountBoth.Au4.text}{zero}
\definevar{codeCountBoth.Au4.1.text}{eight}
\definevar{codeCountBoth.Au4.2.text}{seven}
\definevar{codeCountBoth.Au4.3.text}{two}
\definevar{codeCountBoth.Au4.4.text}{two}
\definevar{codeCountBoth.L1.text}{zero}
\definevar{codeCountBoth.L1.1.text}{zero}
\definevar{codeCountBoth.L1.2.text}{five}
\definevar{codeCountBoth.L1.3.text}{four}
\definevar{codeCountBoth.L2.text}{two}
\definevar{codeCountBoth.L2.1.text}{four}
\definevar{codeCountBoth.L2.2.text}{zero}
\definevar{codeCountBoth.L2.3.text}{zero}
\definevar{codeCountBoth.L3.text}{one}
\definevar{codeCountBoth.L3.1.text}{two}
\definevar{codeCountBoth.L3.2.text}{three}
\definevar{codeCountBoth.L3.3.text}{nine}
\definevar{codeCountBoth.I1.text}{zero}
\definevar{codeCountBoth.I1.1.text}{seven}
\definevar{codeCountBoth.I1.2.text}{zero}
\definevar{codeCountBoth.I1.3.text}{zero}
\definevar{codeCountBoth.I1.4.text}{zero}
\definevar{codeCountBoth.I1.5.text}{one}
\definevar{codeCountBoth.I2.text}{zero}
\definevar{codeCountBoth.I2.1.text}{two}
\definevar{codeCountBoth.I2.2.text}{12}
\definevar{codeCountBoth.I2.3.text}{zero}
\definevar{codeCountBoth.I2.4.text}{zero}
\definevar{codeCountBoth.I2.5.text}{two}
\definevar{codeCountBoth.I2.6.text}{four}
\definevar{codeCountBoth.I3.text}{zero}
\definevar{codeCountBoth.I3.1.text}{one}
\definevar{codeCountBoth.I3.2.text}{one}
\definevar{codeCountBoth.I3.3.text}{17}
\definevar{codeCountBoth.I3.4.text}{one}
\definevar{codeCountBoth.I4.text}{zero}
\definevar{codeCountBoth.I4.1.text}{zero}
\definevar{codeCountBoth.I4.2.text}{zero}
\definevar{codeCountBoth.I4.3.text}{zero}
\definevar{codeCountBoth.I4.4.text}{two}
\definevar{codeCountBoth.I4.5.text}{one}
\definevar{codeCountBoth.I4.6.text}{zero}
\definevar{codeCountBoth.I4.7.text}{one}
\definevar{codeCountBoth.I4.8.text}{eight}
\definevar{codeCountBoth.I4.9.text}{three}
\definevar{codeCountBoth.I4.10.text}{one}
\definevar{codeCountBoth.I4.11.text}{zero}
\definevar{codeCountBoth.I4.12.text}{one}
\definevar{codeCountBoth.I4.13.text}{one}
\definevar{codeCountBoth.I4.14.text}{seven}
\definevar{codeCountBoth.I5.text}{zero}
\definevar{codeCountBoth.I5.1.text}{one}
\definevar{codeCountBoth.I5.2.text}{zero}
\definevar{codeCountBoth.I6.text}{zero}
\definevar{codeCountBoth.I6.1.text}{zero}
\definevar{codeCountBoth.I6.2.text}{zero}
\definevar{codeCountBoth.I6.3.text}{zero}
\definevar{codeCountBoth.I6.4.text}{zero}
\definevar{codeCountBoth.I6.5.text}{zero}
\definevar{codeCountBoth.I6.6.text}{zero}
\definevar{codeCountBoth.I6.7.text}{zero}
\definevar{codeCountBoth.I6.8.text}{zero}
\definevar{codeCountBoth.I6.9.text}{zero}
\definevar{codeCountBoth.I6.10.text}{zero}
\definevar{codeCountBoth.I6.11.text}{zero}
\definevar{codeCountBoth.I6.12.text}{zero}
\definevar{codeCountBoth.I7.text}{zero}
\definevar{codeCountBoth.I7.1.text}{zero}
\definevar{codeCountBoth.I7.2.text}{zero}
\definevar{codeCountBoth.I7.3.text}{zero}
\definevar{codeCountBoth.I7.4.text}{zero}
\definevar{codeCountBoth.I7.5.text}{zero}
\definevar{codeCountBoth.I7.6.text}{zero}
\definevar{codeCountBoth.I7.7.text}{two}
\definevar{codeCountBoth.I7.8.text}{one}
\definevar{codeCountBoth.I8.text}{zero}
\definevar{codeCountBoth.I8.1.text}{zero}
\definevar{codeCountBoth.I8.2.text}{zero}
\definevar{codeCountBoth.I8.3.text}{zero}
\definevar{codeCountBoth.I9.text}{zero}
\definevar{codeCountBoth.I9.1.text}{one}
\definevar{codeCountBoth.I9.2.text}{zero}
\definevar{codeCountBoth.I9.3.text}{zero}
\definevar{codeCountBoth.I9.4.text}{zero}
\definevar{codeCountBoth.I9.5.text}{zero}
\definevar{codeCountBoth.I9.6.text}{zero}
\definevar{codeCountBoth.I9.7.text}{zero}
\definevar{codeCountBoth.I9.8.text}{one}
\definevar{codeCountBoth.I9.9.text}{zero}
\definevar{codeCountBoth.I9.10.text}{zero}
\definevar{codeCountBoth.I10.text}{zero}
\definevar{codeCountBoth.I10.1.text}{zero}
\definevar{codeCountBoth.I10.2.text}{zero}
\definevar{codeCountBoth.I10.3.text}{zero}
\definevar{codeCountBoth.I10.4.text}{zero}
\definevar{codeCountBoth.I10.5.text}{one}
\definevar{codeCountBoth.I10.6.text}{one}
\definevar{codeCountBoth.I10.7.text}{zero}
\definevar{codeCountBoth.I10.8.text}{zero}
\definevar{codeCountBoth.I10.9.text}{zero}
\definevar{codeCountBoth.I10.10.text}{zero}
\definevar{codeCountBoth.M1.text}{zero}
\definevar{codeCountBoth.M1.1.text}{10}
\definevar{codeCountBoth.M1.2.text}{five}
\definevar{codeCountBoth.M1.3.text}{five}
\definevar{codeCountBoth.M1.4.text}{seven}
\definevar{codeCountBoth.M2.text}{zero}
\definevar{codeCountBoth.M2.1.text}{six}
\definevar{codeCountBoth.M2.2.text}{two}
\definevar{codeCountBoth.M3.text}{zero}
\definevar{codeCountBoth.M3.1.text}{five}
\definevar{codeCountBoth.M4.text}{zero}
\definevar{codeCountBoth.M4.1.text}{one}
\definevar{codeCountBoth.M4.2.text}{one}
\definevar{codeCountBoth.M4.3.text}{15}
\definevar{codeCountBoth.M4.4.text}{one}
\definevar{codeCountBoth.M4.5.text}{nine}
\definevar{codeCountBoth.M4.6.text}{four}
\definevar{codeCountBoth.M4.7.text}{one}
\definevar{codeCountBoth.M4.8.text}{one}
\definevar{codeCountBoth.M4.9.text}{six}
\definevar{codeCountBoth.M4.10.text}{four}
\definevar{codeCountBoth.M4.11.text}{two}
\definevar{codeCountBoth.M4.12.text}{one}
\definevar{codeCountBoth.M4.13.text}{seven}
\definevar{codeCountBoth.M4.14.text}{two}
\definevar{codeCountBoth.M4.15.text}{two}
\definevar{codeCountBoth.M4.16.text}{one}
\definevar{codeCountBoth.M4.17.text}{one}
\definevar{codeCountBoth.E1.text}{zero}
\definevar{codeCountBoth.E1.1.text}{one}
\definevar{codeCountBoth.E1.2.text}{one}
\definevar{codeCountBoth.E1.3.text}{two}
\definevar{codeCountBoth.E2.text}{zero}
\definevar{codeCountBoth.E2.1.text}{three}
\definevar{codeCountBoth.E2.2.text}{two}
\definevar{codeCountBoth.E2.3.text}{one}
\definevar{codeCountBoth.E2.4.text}{one}
\definevar{codeCountBoth.E2.5.text}{19}
\definevar{codeCountBoth.E2.6.text}{zero}
\definevar{codeCountBoth.E2.7.text}{one}
\definevar{codeCountBoth.E2.8.text}{zero}
\definevar{codeCountBoth.E2.9.text}{zero}
\definevar{codeCountBoth.E2.10.text}{six}
\definevar{codeCountBoth.E2.11.text}{two}
\definevar{codeCountBoth.E2.12.text}{one}
\definevar{codeCountBoth.E3.text}{zero}
\definevar{codeCountBoth.E3.1.text}{three}
\definevar{codeCountBoth.E3.2.text}{zero}
\definevar{codeCountBoth.E3.3.text}{10}
\definevar{codeCountBoth.E3.4.text}{zero}
\definevar{codeCountBoth.E4.text}{zero}
\definevar{codeCountBoth.E4.1.text}{11}
\definevar{codeCountBoth.E4.2.text}{zero}
\definevar{codeCountBoth.E4.3.text}{five}
\definevar{codeCountBoth.E4.4.text}{zero}
\definevar{codeCountBoth.E4.5.text}{one}
\definevar{codeCountBoth.E4.6.text}{zero}
\definevar{codeCountBoth.E5.text}{zero}
\definevar{codeCountBoth.E5.1.text}{zero}
\definevar{codeCountBoth.E5.2.text}{zero}
\definevar{codeCountBoth.E5.3.text}{zero}
\definevar{codeCountBoth.E5.4.text}{zero}
\definevar{codeCountBoth.E5.5.text}{zero}
\definevar{codeCountBoth.E5.6.text}{one}
\definevar{codeCountBoth.E6.text}{zero}
\definevar{codeCountBoth.E6.1.text}{zero}
\definevar{codeCountBoth.E6.2.text}{zero}
\definevar{codeCountBoth.E6.3.text}{zero}
\definevar{codeCountBoth.E6.4.text}{zero}
\definevar{codeCountBoth.E6.5.text}{zero}
\definevar{codeCountBoth.E6.6.text}{zero}
\definevar{codeCountBoth.E6.7.text}{zero}
\definevar{codeCountBoth.E6.8.text}{one}
\definevar{codeCountBoth.E7.text}{zero}
\definevar{codeCountBoth.E7.1.text}{zero}
\definevar{codeCountBoth.E7.2.text}{zero}
\definevar{codeCountBoth.E7.3.text}{zero}
\definevar{codeCountBoth.E7.4.text}{zero}
\definevar{codeCountBoth.E7.5.text}{zero}
\definevar{codeCountBoth.Soc1.text}{zero}
\definevar{codeCountBoth.Soc1.1.text}{five}
\definevar{codeCountBoth.Soc1.2.text}{six}
\definevar{codeCountBoth.Soc1.3.text}{zero}
\definevar{codeCountBoth.Soc1.4.text}{one}
\definevar{codeCountBoth.Soc1.5.text}{seven}
\definevar{codeCountBoth.Soc1.6.text}{zero}
\definevar{codeCountBoth.Soc1.7.text}{four}
\definevar{codeCountBoth.Soc1.8.text}{four}
\definevar{codeCountBoth.Soc1.9.text}{one}
\definevar{codeCountBoth.Soc1.10.text}{two}
\definevar{codeCountBoth.Soc1.11.text}{four}
\definevar{codeCountBoth.Soc1.12.text}{one}
\definevar{codeCountBoth.Soc1.13.text}{zero}
\definevar{codeCountBoth.Soc2.text}{zero}
\definevar{codeCountBoth.Soc2.1.text}{zero}
\definevar{codeCountBoth.Soc2.2.text}{zero}
\definevar{codeCountBoth.Soc2.3.text}{zero}
\definevar{codeCountBoth.Soc2.4.text}{zero}
\definevar{codeCountBoth.Soc2.5.text}{zero}
\definevar{codeCountBoth.Soc2.6.text}{three}
\definevar{codeCountBoth.Soc3.text}{zero}
\definevar{codeCountBoth.Soc3.1.text}{12}
\definevar{codeCountBoth.Soc3.2.text}{one}
\definevar{codeCountBoth.Soc3.3.text}{seven}
\definevar{codeCountBoth.Soc3.4.text}{four}
\definevar{codeCountBoth.Soc3.5.text}{four}
\definevar{codeCountBoth.Soc4.text}{zero}
\definevar{codeCountBoth.Soc4.1.text}{one}
\definevar{codeCountBoth.C1.text}{zero}
\definevar{codeCountBoth.C1.1.text}{zero}
\definevar{codeCountBoth.C1.2.text}{zero}
\definevar{codeCountBoth.C1.3.text}{zero}
\definevar{codeCountBoth.C2.text}{zero}
\definevar{codeCountBoth.C2.1.text}{zero}
\definevar{codeCountBoth.C2.2.text}{zero}
\definevar{codeCountBoth.C3.text}{zero}
\definevar{codeCountBoth.C3.1.text}{zero}
\definevar{codeCountBoth.C3.2.text}{zero}
\definevar{codeCountBoth.A1.text}{zero}
\definevar{codeCountBoth.A1.1.text}{10}
\definevar{codeCountBoth.A1.2.text}{five}
\definevar{codeCountBoth.A1.3.text}{zero}
\definevar{codeCountBoth.A1.4.text}{zero}
\definevar{codeCountBoth.A1.5.text}{zero}
\definevar{codeCountBoth.A1.6.text}{zero}
\definevar{codeCountBoth.A1.7.text}{zero}
\definevar{codeCountBoth.A1.8.text}{one}
\definevar{codeCountBoth.A1.9.text}{two}
\definevar{codeCountBoth.A1.10.text}{one}
\definevar{codeCountBoth.A1.11.text}{three}
\definevar{codeCountBoth.A1.12.text}{one}
\definevar{codeCountBoth.A1.13.text}{one}
\definevar{codeCountBoth.A1.14.text}{four}
\definevar{codeCountBoth.A2.text}{zero}
\definevar{codeCountBoth.A2.1.text}{11}
\definevar{codeCountBoth.A2.2.text}{three}
\definevar{codeCountBoth.A2.3.text}{zero}
\definevar{codeCountBoth.A2.4.text}{zero}
\definevar{codeCountBoth.A2.5.text}{zero}
\definevar{codeCountBoth.A2.6.text}{one}
\definevar{codeCountBoth.A2.7.text}{four}
\definevar{codeCountBoth.A2.8.text}{two}
\definevar{codeCountBoth.A2.9.text}{zero}
\definevar{codeCountBoth.A3.text}{zero}
\definevar{codeCountBoth.A3.1.text}{24}
\definevar{codeCountBoth.A3.2.text}{zero}
\definevar{codeCountBoth.A3.3.text}{zero}
\definevar{codeCountBoth.A3.4.text}{zero}
\definevar{codeCountBoth.A3.5.text}{zero}
\definevar{codeCountBoth.Sec1.text}{zero}
\definevar{codeCountBoth.Sec1.1.text}{eight}
\definevar{codeCountBoth.Sec1.2.text}{zero}
\definevar{codeCountBoth.Sec1.3.text}{zero}
\definevar{codeCountBoth.Sec1.4.text}{one}
\definevar{codeCountBoth.Sec1.5.text}{11}
\definevar{codeCountBoth.Sec1.6.text}{one}
\definevar{codeCountBoth.Sec1.7.text}{one}
\definevar{codeCountBoth.Sec1.8.text}{three}
\definevar{codeCountBoth.Sec1.9.text}{one}
\definevar{codeCountBoth.Sec1.10.text}{one}
\definevar{codeCountBoth.Sec1.11.text}{two}
\definevar{codeCountBoth.Sec1.12.text}{three}
\definevar{codeCountBoth.Sec1.13.text}{two}
\definevar{codeCountBoth.Sec1.14.text}{two}
\definevar{codeCountBoth.Sec2.text}{zero}
\definevar{codeCountBoth.Sec2.1.text}{34}
\definevar{codeCountBoth.Sec2.2.text}{five}
\definevar{codeCountBoth.Sec2.3.text}{one}
\definevar{codeCountBoth.Sec2.4.text}{three}
\definevar{codeCountBoth.Sec2.5.text}{one}
\definevar{codeCountBoth.DS1.text}{zero}
\definevar{codeCountBoth.DS1.1.text}{two}
\definevar{codeCountBoth.DS1.2.text}{one}
\definevar{codeCountBoth.DS2.text}{zero}
\definevar{codeCountBoth.DS2.1.text}{zero}
\definevar{codeCountBoth.DS2.2.text}{zero}
\definevar{codeCountBoth.DS2.3.text}{one}
\definevar{codeCountBoth.DS2.4.text}{zero}
\definevar{codeCountBoth.DS2.5.text}{zero}
\definevar{codeCountBoth.DS2.6.text}{zero}
\definevar{codeCountBoth.DS2.7.text}{zero}
\definevar{codeCountBoth.DS2.8.text}{zero}
\definevar{codeCountBoth.DS2.9.text}{two}
\definevar{codeCountBoth.DS2.10.text}{two}
\definevar{codeCountBoth.DS2.11.text}{one}
\definevar{codeCountBoth.DS3.text}{zero}
\definevar{codeCountBoth.DS3.1.text}{zero}
\definevar{codeCountBoth.DS3.2.text}{zero}
\definevar{codeCountBoth.DS3.3.text}{three}
\definevar{codeCountBoth.DS3.4.text}{one}
\definevar{codeCountBoth.DS3.5.text}{zero}
\definevar{codeCountBoth.DS3.6.text}{two}
\definevar{codeCountBoth.DS3.7.text}{one}
\definevar{codeCountBoth.DS3.8.text}{zero}
\definevar{codeCountBoth.DS3.9.text}{zero}
\definevar{codeCountBoth.DS3.10.text}{zero}
\definevar{codeCountBoth.DS3.11.text}{zero}
\definevar{codeCountBoth.DS3.12.text}{one}
\definevar{codeCountBoth.DS3.13.text}{zero}
\definevar{codeCountBoth.DS3.14.text}{zero}
\definevar{codeCountBoth.DS3.15.text}{three}
\definevar{codeCountBoth.DS3.16.text}{three}
\definevar{codeCountBoth.DS3.17.text}{zero}
\definevar{codeCountBoth.DS3.18.text}{11}
\definevar{codeCountBoth.DS3.19.text}{one}
\definevar{codeCountBoth.DS3.20.text}{one}
\definevar{codeCountBoth.DS3.21.text}{two}
\definevar{codeCountBoth.DS3.22.text}{three}
\definevar{codeCountBoth.DS3.23.text}{three}
\definevar{codeCountBoth.DS3.24.text}{one}
\definevar{codeCountBoth.DS3.25.text}{one}
\definevar{codeCountBoth.DS3.26.text}{one}
\definevar{codeCountBoth.DS4.text}{zero}
\definevar{codeCountBoth.DS4.1.text}{zero}
\definevar{codeCountBoth.DS4.2.text}{zero}
\definevar{codeCountBoth.DS4.3.text}{zero}
\definevar{codeCountBoth.DS5.text}{zero}
\definevar{codeCountBoth.DS5.1.text}{one}
\definevar{codeCountBoth.DS5.2.text}{zero}
\definevar{codeCountBoth.DS5.3.text}{zero}
\definevar{codeCountBoth.DS5.4.text}{zero}
\definevar{codeCountBoth.DS5.5.text}{two}
\definevar{codeCountBoth.DS5.6.text}{10}
\definevar{codeCountBoth.DS5.7.text}{zero}
\definevar{codeCountBoth.DS5.8.text}{one}
\definevar{codeCountBoth.DS5.9.text}{one}
\definevar{codeCountBoth.DS5.10.text}{one}
\definevar{codeCountBoth.DS5.11.text}{one}
\definevar{codeCountBoth.DS5.12.text}{two}
\definevar{codeCountBoth.DS5.13.text}{one}
\definevar{codeCountBoth.DS5.14.text}{one}
\definevar{codeCountBoth.DS5.15.text}{six}
\definevar{codeCountBoth.DS5.16.text}{four}
\definevar{codeCountBoth.DS5.17.text}{zero}
\definevar{codeCountBoth.DS5.18.text}{one}
\definevar{codeCountBoth.DS5.19.text}{two}
\definevar{codeCountBoth.DS6.text}{zero}
\definevar{codeCountBoth.DS6.1.text}{zero}
\definevar{codeCountBoth.DS6.2.text}{zero}
\definevar{codeCountBoth.DS6.3.text}{zero}
\definevar{codeCountBoth.Sto1.text}{zero}
\definevar{codeCountBoth.Sto1.1.text}{one}
\definevar{codeCountBoth.Sto1.2.text}{22}
\definevar{codeCountBoth.Sto1.3.text}{four}
\definevar{codeCountBoth.Sto1.4.text}{13}
\definevar{codeCountBoth.Sto1.5.text}{two}
\definevar{codeCountBoth.Sto1.6.text}{one}
\definevar{codeCountBoth.Sto1.7.text}{two}
\definevar{codeCountBoth.Sto1.8.text}{one}
\definevar{codeCountBoth.Sto1.9.text}{six}
\definevar{codeCountBoth.Sto1.10.text}{one}
\definevar{codeCountBoth.Sto2.text}{zero}
\definevar{codeCountBoth.Sto2.1.text}{one}
\definevar{codeCountBoth.Sto2.2.text}{zero}
\definevar{codeCountBoth.Sto2.3.text}{zero}
\definevar{codeCountBoth.Sto2.4.text}{zero}
\definevar{codeCountBoth.Sto2.5.text}{zero}
\definevar{codeCountBoth.Sto2.6.text}{one}
\definevar{codeCountBoth.Sto2.7.text}{one}
\definevar{codeCountBoth.Sto3.text}{zero}
\definevar{codeCountBoth.Sto3.1.text}{two}
\definevar{codeCountBoth.Sto3.2.text}{zero}
\definevar{codeCountBoth.Sto3.3.text}{four}
\definevar{codeCountBoth.Sto3.4.text}{five}
\definevar{codeCountBoth.Sto3.5.text}{two}
\definevar{codeCountBoth.Sto3.6.text}{zero}
\definevar{codeCountBoth.Sto3.7.text}{zero}
\definevar{codeCountBoth.Sto3.8.text}{zero}
\definevar{codeCountBoth.Sto3.9.text}{one}
\definevar{codeCountBoth.Sto4.text}{zero}
\definevar{codeCountBoth.Sto4.1.text}{zero}
\definevar{codeCountBoth.Sto4.2.text}{four}
\definevar{codeCountBoth.Sto4.3.text}{zero}
\definevar{codeCountBoth.Sto4.4.text}{zero}
\definevar{codeCountBoth.Sto5.text}{zero}
\definevar{codeCountBoth.Sto5.1.text}{two}
\definevar{codeCountBoth.Sto5.2.text}{four}
\definevar{codeCountBoth.Sto5.3.text}{one}
\definevar{codeCountBoth.Sto5.4.text}{four}
\definevar{codeCountBoth.Sto5.5.text}{one}
\definevar{codeCountBoth.Sto5.6.text}{seven}
\definevar{codeCountBoth.B1.text}{zero}
\definevar{codeCountBoth.B1.1.text}{zero}
\definevar{codeCountBoth.B1.2.text}{zero}
\definevar{codeCountBoth.B1.3.text}{zero}
\definevar{codeCountBoth.B2.text}{zero}
\definevar{codeCountBoth.B2.1.text}{zero}
\definevar{codeCountBoth.B2.2.text}{zero}
\definevar{codeCountBoth.B2.3.text}{zero}
\definevar{codeCountBoth.B2.4.text}{zero}
\definevar{codeCountBoth.B2.5.text}{zero}
\definevar{codeCountBoth.B2.6.text}{one}
\definevar{codeCountBoth.B2.7.text}{zero}
\definevar{codeCountBoth.B3.text}{zero}
\definevar{codeCountBoth.B3.1.text}{one}
\definevar{codeCountBoth.B3.2.text}{zero}
\definevar{codeCountBoth.B3.3.text}{two}
\definevar{codeCountBoth.B3.4.text}{zero}
\definevar{codeCountBoth.B3.5.text}{zero}
\definevar{codeCountBoth.B3.6.text}{zero}
\definevar{codeCountBoth.B3.7.text}{zero}
\definevar{codeCountBoth.T1.text}{221}
\definevar{codeCountBoth.T2.text}{135}
\definevar{codeCountBoth.Co1.text}{zero}
\definevar{codeCountBoth.Co1.1.text}{two}
\definevar{codeCountBoth.Re1.text}{one}
\definevar{codeCountBoth.Re1.1.text}{two}
\definevar{codeCountBoth.Re1.2.text}{six}
\definevar{codeCountBoth.Re1.3.text}{one}
\definevar{codeCountBoth.Re1.4.text}{one}
\definevar{codeCountBoth.S.text}{zero}
\definevar{codeCountBoth.S1.text}{51}
\definevar{codeCountBoth.S1.1.text}{zero}
\definevar{codeCountBoth.S1.2.text}{zero}
\definevar{codeCountBoth.S1.3.text}{six}
\definevar{codeCountBoth.S1.4.text}{two}
\definevar{codeCountBoth.S1.5.text}{zero}
\definevar{codeCountBoth.S2.text}{37}
\definevar{codeCountBoth.S2.1.text}{two}
\definevar{codeCountBoth.S2.2.text}{zero}
\definevar{codeCountBoth.S2.3.text}{eight}
\definevar{codeCountBoth.S2.4.text}{three}
\definevar{codeCountBoth.S2.5.text}{one}
\definevar{codeCountBoth.S3.text}{zero}
\definevar{codeCountBoth.S3.1.text}{zero}
\definevar{codeCountBoth.S3.2.text}{zero}
\definevar{codeCountBoth.S3.3.text}{zero}
\definevar{codeCountBoth.S3.4.text}{zero}
\definevar{codeCountBoth.S3.5.text}{zero}
\definevar{codeCountBoth.S4.text}{56}
\definevar{codeCountBoth.S4.1.text}{three}
\definevar{codeCountBoth.S4.2.text}{zero}
\definevar{codeCountBoth.S4.3.text}{zero}
\definevar{codeCountBoth.S4.4.text}{zero}
\definevar{codeCountBoth.S4.5.text}{zero}
\definevar{codeCountBoth.S5.text}{35}
\definevar{codeCountBoth.S5.1.text}{14}
\definevar{codeCountBoth.S5.2.text}{zero}
\definevar{codeCountBoth.S5.3.text}{three}
\definevar{codeCountBoth.S5.4.text}{13}
\definevar{codeCountBoth.S5.5.text}{five}
\definevar{codeCountBoth.S6.text}{69}
\definevar{codeCountBoth.S6.1.text}{18}
\definevar{codeCountBoth.S6.2.text}{five}
\definevar{codeCountBoth.S6.3.text}{one}
\definevar{codeCountBoth.S6.4.text}{20}
\definevar{codeCountBoth.S6.5.text}{22}

%% file: numbers/numbers.tex
\definevar{coders}{7}
\definevar{coders.text}{seven}

\definevar{affinity.coders}{5}
\definevar{affinity.coders.text}{five}
\definevar{affinity.categories}{7}
\definevar{affinity.categories.text}{seven}
\definevar{affinity.subcategories}{21}
\definevar{affinity.subcategories.text}{twenty one}

\definevar{unique_codes}{451}
\definevar{content_codes}{444} % without targets+sources
\definevar{used_content_codes}{221}

%% file: sections/01-abstract.tex
The Russian Invasion of Ukraine in early 2022 resulted in a rapidly changing (cyber) threat environment.
This changing environment incentivized the sharing of security advice on social media, both for the Ukrainian population, as well as against Russian cyber attacks at large.  
Previous research found a significant influence of online security advice on end users.

We collected \var{tweets.total} tweets posted after the Russian Invasion of Ukraine and examined \var{tweets.relevant} in detail, including qualitatively coding \var{tweets.coded} relevant tweets and \var{resources.coded} linked documents for security and privacy advice. 
We identified \var{used_content_codes} unique pieces of advice which we divided into seven categories and 21 subcategories, and advice targeted at individuals or organizations. We then compared our findings to those of prior studies, finding noteworthy similarities.
Our results confirm a lack of advice prioritization found by prior work, which seems especially detrimental during times of crisis. In addition, we find offers for individual support to be a valuable tool and identify misinformation as a rising threat in general and for security advice specifically.

%% file: sections/02-intro.tex
In the early hours of 24$^\text{th}$ February 2022, Russian President Vladimir Putin announced a ``special military operation'', launching a large-scale military invasion of neighboring Ukraine.
Following cruise and ballistic missile strikes directed at Ukrainian airfields, military headquarters, and military depots,
Russian troops entered Ukraine from four main directions: north from Belarus, northeast from Russia, east from the \gls{dpr} and \gls{lpr}, and south from the annexed region of Crimea.
Russia’s invasion follows the 2014 annexation of the Crimean peninsula, which was itself followed by eight years of support for separatist rebels in eastern Ukraine.

This change in the global threat environment was accompanied by a sudden change in the cyber threat environment: the invasion was preceded and accompanied by intensified cyber attacks, some of which had physical consequences. For instance, in early February malware targeting several Ukrainian Government and IT organizations was detected and a \gls{ddos} attack made multiple government and banking websites inaccessible for hours. These attacks impacted Ukraine's energy, media, financial, and business sectors. Furthermore, several cyber threats and attacks continued after the invasion as well, impacting the distribution of food, medicines, and supplies. During this period, other malicious activities like disinformation through deep fake technology, phishing emails, use of surveillance software and data-wiper malware, etc., were detected~\cite{jakub2022:attacktimeline}.

The \gls{event-short} resulted in a heightened threat environment for companies outside of Russia and Ukraine, as well.
This change in threat level was also highlighted by a number of advisories by national agencies, including from the US, UK, Germany, Canada, and Australia~\cite{cisa:2022:shieldsup,Acsc:2022:Advisory,ncsc:2022:encourage,bsi:2022:statement,can:2022:bulletin}.
The advisories warn that Russian state-sponsored threat actors and aligned cybercrime groups might be targeting critical industries and organizations in the United States and other Western nations.

One example of an attack such adversaries carried out in the past is the SolarWinds incident which was one of the most detrimental attacks on the supply chain network. It was found that \gls{apt} actors created a backdoor hidden in a software update of SolarWinds' Orion system of.
This attack affected almost 18,000 customers worldwide~\cite{alkhadra2021solarwinds}.
The {NotPetya} attack in 2017 which had an impact globally is another example of a supply chain attack.
A backdoor was planted in an accounting software mostly used by Ukrainian accounting firms~\cite{crosignani2021:SCattack}. 

Following these incidents and attacks, many instances of security-related advice were shared on news sources and social networks like Twitter.
In some cases, security-related advice and information online were conflated with misinformation.
For instance, rumors about Signal, an instant messaging platform, being hacked were widespread.
This was in fact dismissed by Signal as a part of a coordinated misinformation campaign to encourage users to use less secure methods for communication~\cite{mehrotra-2022:signalfakenews}.

In this paper, we examine the security and privacy advice provided around Russia's 2022 Invasion of Ukraine on the social media platform Twitter.
This is especially relevant given the connection of cyber attacks, the success of their mitigation, and global physical consequences of the \gls{event-short}.
We base our research approach on the following research questions:

\boldparagraph{RQ1}{\itshape ``What security and privacy advice was shared on Twitter related to the \gls{event}?''}
We are interested in what security and privacy advice was shared on Twitter between February and May 2022, especially in relation to the heightened cyber threat resulting from the \gls{event-short}.
We  are interested in the tweets as well as resources provided such as linked documents and websites, and the targets of the advice such as companies or individuals, including those in Ukraine as well as other directly or indirectly affected people.

\boldparagraph{RQ2}{\itshape ``How does the advice compare to security and privacy advice shared in other contexts?''}
We explore if the advice around the \gls{event-short} resembles or differs from security and privacy advice collected at different times and in different contexts. To this end, we compare our data to that of previous studies. Additionally, we investigate the relationship between advice and its frequency in our data, and evaluation and prioritization of advice in prior work. As far as possible, we seek to understand if and how the advice was tailored to the situation at hand.
\vspace{4pt plus 2pt minus 1pt}

For this, we create a taxonomy of \var{used_content_codes} pieces of advice during the \gls{event-short}. In seven main categories, we find a wide range of advice including messaging \& social media, organizational policies, and meta advice on sharing security advice. The majority of advice was rather generic, leading to a low correlation with \citeauthor{Redmiles:2020:Comprehensive}'s user priority ranking, which was based on actionability among other criteria~\cite{Redmiles:2020:Comprehensive}. By contrast, we find some significant correlation with their expert priority ranking, as well as advice frequency in the advice for non-tech-savvy users collected by \citeauthor{Reeder:2017:Steps}~\cite{Reeder:2017:Steps}.

This work is structured as follows:
After this general introduction (Section~\ref{sec:intro}),
we discuss related work in the areas of security perceptions \& behavior, social media \& information sharing in crises, as well as security \& privacy advice (Section~\ref{sec:related-work}).
We describe our approach (Section~\ref{sec:method}) and highlight the findings (Section~\ref{sec:results}).
Finally, we discuss our findings (Section~\ref{sec:discussion}) and draw a conclusion (Section~\ref{sec:conclusion}).

%% file: sections/04-related-work.tex
We present and discuss previous work in three related fields:
investigations into security-related user perceptions and behavior,
research involving content on social media and information sharing in crises,
and security or privacy advice for users.
In this section, we also put our work into context and highlight some novel contributions of our research.

\boldparagraph{Influences on Security Behavior}
Prior work has investigated how security behavior is influenced both in general and in vulnerable populations.
Previous studies established connections between user behavior and the user’s perception of risk~\cite{Aytes:2003:Research,Beris:2015:Employee,Redmiles:2018:Dancing}, 
(security) fatigue~\cite{Stanton:2016:Security}, and social influence effects~\cite{Das:2014:Effect,Das:2014:Increasing,Das:2015:Role,Das:2019:Typology}.
% ID15 - lit review, factors that influence security decisions
\citeauthor{Howe:2012:Psychology} performed a literature review of studies investigating factors that influence security decisions, finding delivery of security measures to users as an important factor~\cite{Howe:2012:Psychology}.
In the context of scams and vulnerable populations, \citeauthor{Vitak:2018:Knew} conducted interviews with 52 families from high-poverty communities, finding a complex relationship between participants’ negative experiences, their perceptions and their general mistrust of sharing data through online channels~\cite{Vitak:2018:Knew}.
Further research into methods for influencing security behavior includes nudges and warnings.
Previous studies include a literature assessment~\cite{Acquisti:2017:Nudges}, experiments reminding users of updates and \gls{2fa}~\cite{Frik:2018:Better}, and security dialog attractors~\cite{Bravo:2013:Your,Busse:2019:Snooze}.
Vulnerable persons and helpers are often the target of scams and phishing attacks during crises.
\citeauthor{Egelman:2008:You} examined in a lab study with 60 participants the effectiveness of phishing warnings, finding that 97\% participants fell for at least one of the phishing messages~\cite{Egelman:2008:You}. 
These prior studies on how security behavior is influenced inform our view on and discussion of the advice we collected.

\boldparagraph{Social Media \& Information Sharing in Crises}

Social media reactions to the events around the 2014 Russian annexation of the Crimea peninsula have been extensively investigated in research~\cite{Ronzhyn:2014:Use,Suslov:2014:Crimea}, specifically, topics~\cite{Mishler:2015:Using}, hashtags~\cite{Makhortykh:2015:Savedonbasspeople}, images~\cite{Pantti:2019:Personalisation}, and memes~\cite{Wiggins:2016:Crimea}.
Twitter and other social media are a common data source for research, including newcomers' experiences~\cite{Burke:2009:Feed}, audience perceptions~\cite{Bernstein:2013:Quantifying}, information sharing~\cite{Christofides:2012:Hey,Syn:2015:Social}, and rumors~\cite{Zubiaga:2018:Detection}.
This includes specific user types such as journalists~\cite{Lee:2015:Double} or government departments~\cite{Depaula:2018:Information}.
\citeauthor{Sit:2019:Identifying} introduce a labeled dataset of disaster-related tweets and  present a series of deep learning and machine learning methods for a binary classification of disaster relatedness~\cite{Sit:2019:Identifying}.
\citeauthor{Imran:2015:Processing} surveyed the state of the art regarding computational methods to process social media messages and highlight both their contributions and shortcomings~\cite{Imran:2015:Processing}.
In the area of crisis research, multiple publications systematize previous work based on social media data~\cite{Wang:2018:Social,Reuter:2018:Fifteen,Reuter:2018:Social}.
Works specifically investigated information aggregation on Reddit~\cite{Leavitt:2017:Role}, and Twitter posts around crises~\cite{Olteanu:2015:Expect} and their comprehension~\cite{Temnikova:2015:Case}.
Specific cases discussed include 2012 Hurricane Sandy~\cite{Leavitt:2014:Upvoting}, the 2013 Gezi Park protests in Turkey~\cite{Ozduzen:2020:Digital}, and the 2015--2016 Zika virus outbreak~\cite{Hagen:2018:Crisis}.
The spread of misinformation during crises was studied in relation to the emotional proximity of users~\cite{Starbird:2015:Connected}, and with regard to Russian influence operations within \#BlackLivesMatter~\cite{Starbird:2018:BLM}.

The \gls{event} went hand-in-hand with surveillance and censorship by the Russian state.
Specifically for Russian internet infrastructure,  prior research considered internet governance~\cite{Kukkola:2018:Civilian,Claessen:2020:Reshaping}, throttling of Twitter domains~\cite{Xue:2021:Throttling}, and deployed traffic filtering solutions~\cite{Ermoshina:2022:Market}.
\citeauthor{Ermoshina:2017:Migrating} analyzed internet censorship-resistance tactics by Russian users, content produces, and service providers~\cite{Ermoshina:2017:Migrating}.
More recently, \citeauthor{Gabdulhakov:2020:Trolling} investigated a wave of social media user arrests in Russia in semi-structured in-depth interviews with lawyers, rights defenders, academics, \glspl{ngo}, and law enforcement authorities~\cite{Gabdulhakov:2020:Trolling}.
\citeauthor{Akbari:2019:Platform} present an analysis of the challenges faced by the Telegram messenger in Russia and Iran~\cite{Akbari:2019:Platform}.

\boldparagraph{Security \& Privacy Advice}

Previous research investigated security advice in the context of experts vs.\ users~\cite{Ion:2015:no,Busse:2019:Replication}, and for older adults~\cite{Nicholson:2019:Older}.
Multiple publications investigated the adoption and impact of security practices~\cite{Fagan:2016:They,Dekoven:2019:Measuring,Zou:2020:Examining}.
Respondents' security advice sources were investigated in interviews~\cite{Redmiles:2016:Think} and surveys~\cite{Rader:2015:Identifying,Redmiles:2016:Learned}, as well as specific advice for developers~\cite{Acar:2017:Developers}.
\citeauthor{Herley:2009:Advice} postulates that by evaluating (security) advice solely on benefit, we have implicitly valued user time and effort at zero~\cite{Herley:2009:Advice}. This  becomes and important aspect in the light of recent studies, which find a large number of advice pieces.

\citeauthor{Tahaei:2022:Understanding} qualitatively analyzed 119 privacy-related accepted
answers on Stack Overflow, extracting 148 pieces of advice~\cite{Tahaei:2022:Understanding}.

\citeauthor{Reeder:2017:Steps} collected 152 pieces of advice by asking security experts for the top three recommendations they would give to non-tech-savvy users~\cite{Reeder:2017:Steps}.
\citeauthor{Redmiles:2020:Comprehensive} conducted a measurement study to identify 374
unique recommended behaviors contained within 1,264 documents of online security and privacy advice and evaluated the security advice in a user-study with 1,586 users and 41 professional security experts~\cite{Redmiles:2020:Comprehensive}.
\citeauthor{Boyd:2021:Understanding} collected 41 safety guides distributed during \gls{blm} protests and surveyed 167 protesters, finding that many were unaware of key advice like using end-to-end encrypted messengers~\cite{Boyd:2021:Understanding}.

We compare our collection of pieces of advice and online documents shared on Twitter during the \gls{event} to these prior studies, and provide novel insight on what security and privacy advice is distributed during crises that directly impact the cyber threat environment.

%% file: sections/05-method.tex
%% Purpose and Approach
In this section, we provide an overview of our methodology for assessing online security and privacy advice related to the \gls{event}, including data collection from \gls{tw} and documents linked on Twitter between  February and May 2022. 
We also detail our qualitative codebook and coding process, highlight our ethical considerations, and discuss the limitations of our work.

\subsection{Study Setup}
To gain insight into security and privacy advice shared around the \gls{event},
we collected and analyzed \var{tweets.total} tweets for their relevance and examined \var{tweets.coded} posts in detail for security and privacy advice.
As we were especially interested in widely shared advice and resources, we decided to study public data on Twitter. Twitter had been successfully used to analyze the spread of information during crises in several prior studies~\cite{Hagen:2018:Crisis, Olteanu:2015:Expect, Schroeder:2012:Arab, Starbird:2018:BLM, Boyd:2021:Understanding}.

\boldparagraph{Data Collection}
We collected security and privacy advice and resources shared on \gls{tw} during the \gls{event} from February to May 2022.
The tweets were collected using the official Twitter API for Academic Research\footnote{https://developer.twitter.com/en/products/twitter-api/academic-research} and \gls{tw} Streams\footnote{https://developer.twitter.com/en/docs/tutorials/stream-tweets-in-real-time} using the Python library Tweepy\footnote{https://github.com/tweepy/tweepy}.
The results were further enhanced by using the unofficial \gls{tw} API and the Python library Twint\footnote{https://github.com/twintproject/twint}, which allows scraping by hashtags.
We used a list of keywords aiming to cover and gather all relevant tweets.
This resulted in \var{tweets.total} tweets that could possibly be relevant.
Following this, we applied a manual filtering process:
In a first round, each tweet was marked as security/privacy advice if at least one of two coders deemed it relevant.
Tweets that did not include or refer to security and privacy advice were discarded.
In a second round, one coder verified if the remaining tweets mentioned Ukraine or the \gls{event} in any way.
A second coder crosschecked >10\% of the tweets that the first coder had categorized as non-related to ensure that no relevant data was missed, finding no additional relevant data points. 
After this manual filtering process, we had \var{tweets.coded} tweets remaining.
From these tweets, we additionally collected any links to external documents, which resulted in a total of \var{resources.coded} documents.
Both the tweets and the documents (denoted with prefixes T and D, respectively) were then analyzed as detailed below and as depicted in Figure~\ref{fig:pipeline}. We make no further distinction between the two and collectively call them resources in our results. 

\boldparagraph{Ethical Considerations \& Data Protection}
Our institutions did not require \gls{irb} approval for this type of public information measurement study.

When working with data during a crisis, ethical considerations are essential to the study design, analysis, and reporting.
Due to the potential of targeted threats from sophisticated attackers,
our focus was on ensuring that reporting in this study would not harm the population as a whole or particular individuals more.
As such, we do not report potentially compromising data. Out of ethical concerns, we decided against contacting people  who live in a war zone or had recently fled one for interviews or other direct interaction to avoid bothering potentially traumatized people and focused on publicly available data instead~\cite{Bhalerao:2022:ethics}. We stored all data protected from unauthorized access by encryption and access control. While all data was public at the time of collection, we refrain from republishing it alongside this work to preserve people's privacy and control over how their identifiable data is shared, as well as their ability to delete their data.

\subsection{Data Analysis}

\input{includes/graph-analysis-pipeline}

We outline our data analysis pipeline below and in Figure~\ref{fig:pipeline}.
Our goal in analyzing the security and privacy advice was to create a taxonomy of the different types of advice shared during the \gls{event-short} and to compare it to types of advice that prior work has found.

For our study with advice and resource artifacts, we evaluated both qualitative and quantitative data points.
We analyzed all collected tweets and documents in an iterative open coding approach~\cite{Charmaz:2014:grounded,Strauss:1997:grounded,Corbin:1990:grounded}.
All researchers together created an initial codebook based on previous work that collected pieces of advice from Twitter data as well as other sources (\cite{Redmiles:2020:Comprehensive,Reeder:2017:Steps,Rader:2015:Identifying,Zou:2020:Examining}).
Tweets and linked documents were then coded by \var{coders.text} coders, resolving conflicts by consensus decision or by introducing new (sub)codes.
This approach does not necessitate the reporting of intercoder agreement, because
each conflict is resolved as it emerges, resulting in a hypothetical final agreement of 100\%~\cite{McDonald:2019:IRR}.
Our final codebook consisted of \var{unique_codes} unique codes. Seven codes served to distinguish sources and targets of advice. Of the \var{content_codes} codes referring to pieces of advice, \var{used_content_codes} were assigned at least once. Unused codes from prior work were kept at count zero for the comparison.

To investigate emerging themes and directions in our codes, we used the affinity diagramming method~\cite{Beyer:1997:affinity} on the codes we had assigned at least once. We conducted a collaborative affinity diagramming session with \var{affinity.coders.text} researchers and iteratively established seven categories and 21 subcategories. An overview is presented in Table \ref{tab:taxonomy}.
Finally, in order to compare our findings with prior work, we manually matched our codebook to theirs. We qualitatively analyzed the top ten corresponding codes from each data set and computed correlation using Spearman's correlation coefficient~\cite{Spearman:1987:correlation}. Frequencies of advice were normalized for the number of resources that were coded in each data set, and advice not present in at least one codebook was omitted.

\subsection{Limitations}
Our work includes a number of limitations typical for this type of measurement study and should be interpreted in context.
Given our method of data collection, it is possible that we have missed some advice or types of advice. 
Even though Twitter data is commonly used during crises around the world~\cite{Hagen:2018:Crisis, Olteanu:2015:Expect, Schroeder:2012:Arab, Starbird:2018:BLM, Boyd:2021:Understanding} and gave us rich insights into advice targeted at those affected by the \gls{event-short}, data obtained from Twitter may not be representative of all available advice sources, meaning that our data set may not fully represent the entirety of advice given in the context of the Ukraine war. To mitigate this risk, we only applied very broad filters to our initial data collection, and thereafter manually coded data points for their relevance. Additionally, we followed links to advice sources outside of Twitter and included these documents in our data set. As none of us speak Ukrainian, we may have encountered a language barrier. 
We did, however, include translated search terms during data collection and received useful results from automated translation tools, which allowed us to code tweets in various languages.
Finally, errors or misunderstandings may have occurred during our manual coding process. 
We minimized this risk by having independently coded each tweet and document by at least two researchers and resolving any emerging conflicts.

%% file: includes/graph-analysis-pipeline.tex
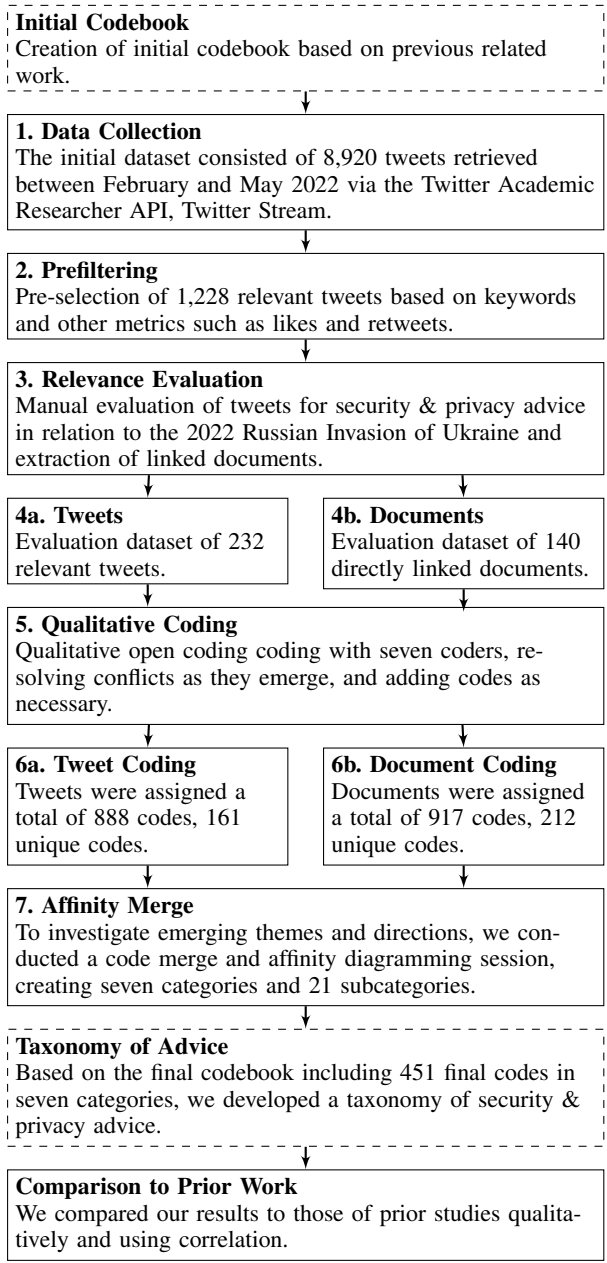
\begin{figure}[tbh!]
    \centering
    %\footnotesize
    \small
    \begin{tikzpicture}[%
        auto,
        % Main
        block_main/.style ={rectangle, draw=black, fill=none, text width=0.9\columnwidth, text ragged, minimum height=3em, inner sep=3pt},
        % Dashed
        block_dash/.style ={rectangle, dashed, draw=black, fill=none, text width=0.9\columnwidth, text ragged, minimum height=3em, inner sep=3pt},
        % Double
        block_double/.style ={rectangle, draw=black, fill=none, text width=0.41\columnwidth, text ragged, minimum height=3em, inner sep=3pt},
        % No Border
        block_noborder/.style ={rectangle, draw=none, fill=none, minimum height=1em, inner sep=0pt, text width=0.90\columnwidth, text centered},
        % Attached
        block_attach/.style ={rectangle, draw=none, fill=none, inner sep=3pt, minimum width=0.9\columnwidth},
        line/.style ={draw, thick, -latex', shorten >=0pt}]
    ]
	\begin{scope}[node distance = 0.3cm]
        \node [block_dash] (codebook) {
            \textbf{Initial Codebook}\\
            Creation of initial codebook based on previous related work.
        };
        \node [block_main, below=of codebook] (collection) {
            \textbf{1. Data Collection}\\
            The initial dataset consisted of \var{tweets.total} tweets retrieved between February and May 2022 via the Twitter Academic Researcher API, Twitter Stream.
        };
        \node [block_main, below=of collection] (prefiltering) {
            \textbf{2. Prefiltering}\\
            Pre-selection of \var{tweets.relevant} relevant tweets based on keywords and other metrics such as likes and retweets.
        };
        \node [block_main, below=of prefiltering] (relevance) {
            \textbf{3. Relevance Evaluation}\\
            Manual evaluation of tweets for security \& privacy advice in relation to the \gls{event} and extraction of linked documents.
   
        };
        \node [block_double, below=of relevance.south west, anchor=north west] (tweets-rel) {
            \textbf{4a. Tweets}\\
            Evaluation dataset of \var{tweets.coded} relevant tweets.
        };
        \node [block_double, below=of relevance.south east, anchor=north east] (resources-rel) {
            \textbf{4b. Documents}\\
            Evaluation dataset of \var{resources.coded} directly linked documents.
        };
        \node [block_main, below=of tweets-rel.south west, anchor=north west] (coding) {
            \textbf{5. Qualitative Coding}\\
            Qualitative open coding coding with \var{coders.text} coders, resolving conflicts as they emerge, and adding codes as necessary.
        };
        \node [block_double, below=of coding.south west, anchor=north west] (tweets-coding) {
            \textbf{6a. Tweet Coding}\\
            Tweets were assigned a total of \var{codes.assigned.tweets} codes, \var{codes.unique.tweets} unique codes.
        };
        \node [block_double, below=of coding.south east, anchor=north east] (resources-coding) {
            \textbf{6b. Document Coding}\\
            Documents were assigned a total of \var{codes.assigned.resources} codes, \var{codes.unique.resources} unique codes.
        };
        \node [block_main, below=of tweets-coding.south west, anchor=north west] (affinity) {
            \textbf{7. Affinity Merge}\\
            To investigate emerging themes and directions, we conducted a code merge and affinity diagramming session, creating seven categories and 21 subcategories.
        };
        \node [block_dash, below=of affinity] (taxonomy) {
            \textbf{Taxonomy of Advice}\\
            Based on the final codebook including \var{unique_codes} final codes in seven categories, we developed a taxonomy of security \& privacy advice.
        };
         \node [block_main, below=of taxonomy] (comparison) {
            \textbf{Comparison to Prior Work}\\
            We compared our results to those of prior studies qualitatively and using correlation.
        };
    % connecting nodes with paths
    \end{scope}
    \begin{scope}[every path/.style=line]
        % Main survey
        \path (codebook) -- (collection);
        \path (collection) -- (prefiltering);
        \path (prefiltering) -- (relevance);
        \path ($(relevance.south west)!0.475!(relevance.south)$) -- (tweets-rel.north);
        \path ($(relevance.south)!0.525!(relevance.south east)$) -- (resources-rel.north);
        \path (tweets-rel.south) --++ (0, -0.3);
        \path (resources-rel.south) --++ (0, -0.3);;
        \path ($(coding.south west)!0.475!(coding.south)$) -- (tweets-coding.north);
        \path ($(coding.south)!0.525!(coding.south east)$) -- (resources-coding.north);
        \path (tweets-coding.south) --++ (0, -0.3);
        \path (resources-coding.south) --++ (0, -0.3);
        \path (affinity) -- (taxonomy);
        \path (taxonomy) -- (comparison);
    \end{scope}
    \end{tikzpicture}
\caption{%
Illustration of the data analysis pipeline. 
Based on the final codes, we created a taxonomy of security \& privacy advice surrounding the \gls{event}, and conducted a comparison with prior work.
}\label{fig:pipeline}
\end{figure}

%% file: sections/06-results.tex
In this section, we present the results of our analysis of the final corpus of \var{tweets.coded} coded tweets and \var{resources.coded} coded documents.
The set of coded tweets has a median number of likes of \var{tweets.likes.median} (sd: \var{tweets.likes.sd}) and a median number of retweets of \var{tweets.retweets.median} (sd: \var{tweets.retweets.sd}). 
We first report on the taxonomy of the advice we created, detailing what advice was shared in connection to the \gls{event-short} by and for whom (Subsection \ref{subsec:analysis}). 
Secondly, we describe the results of comparing our data to previously collected security and privacy advice and its evaluation (Subsection \ref{subsec:comparison}). 
While we do give indications how many resources contained advice, our findings are qualitative in nature.

\subsection{Analysis of Advice}\label{subsec:analysis}

In our analysis, we identified five types of advice sources and distinguished between advice targeted at individuals and organizations. We present an overview in Table~\ref{table:resource_source_target}.

\begin{figure}[h]
    \centering
    \includegraphics[width=\columnwidth]{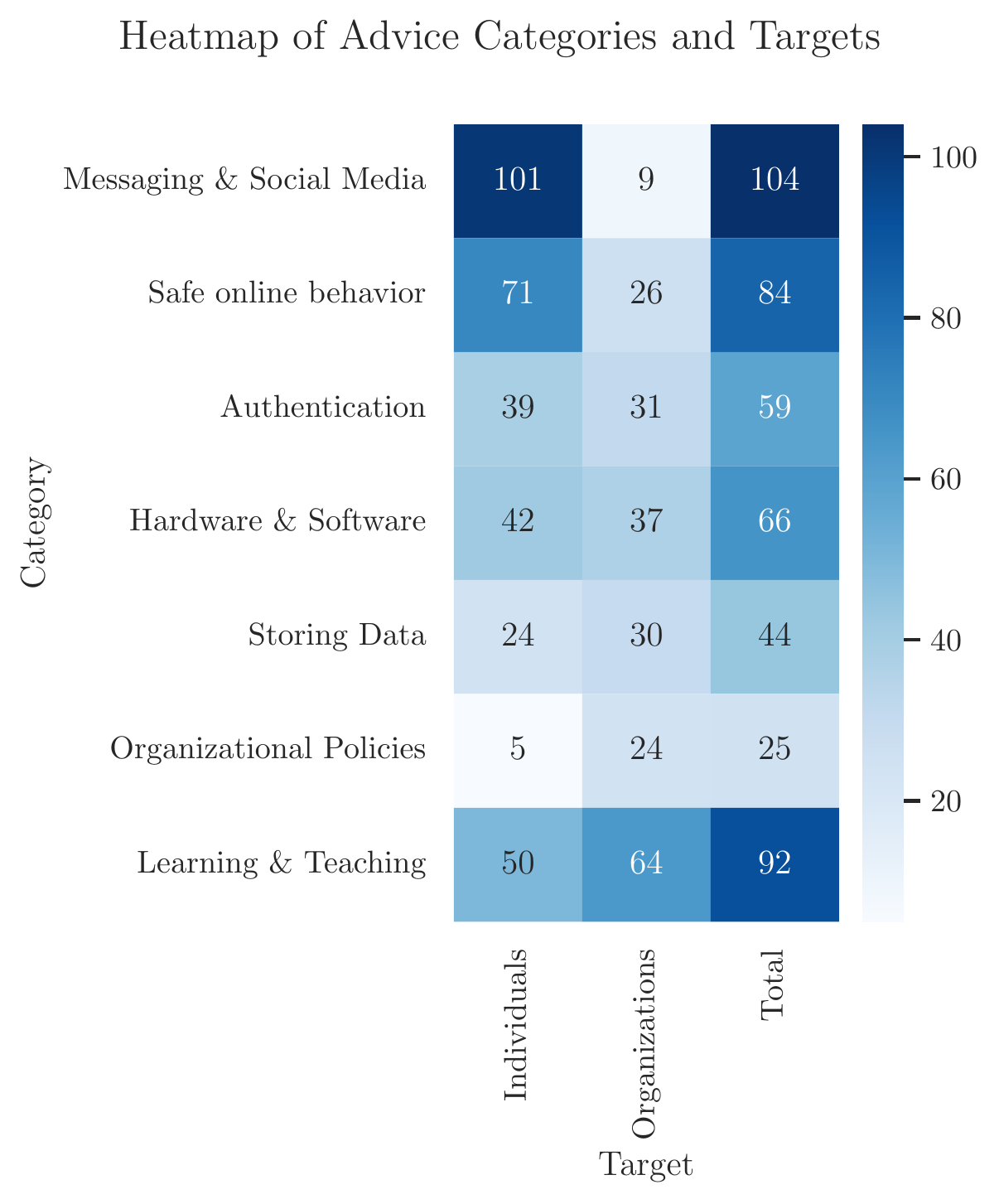}
    \caption{Heatmap of code categories and the number of times they appeared in advice resources, differentiated by target group.}
    \label{fig:heatmap}
\end{figure}

Below, we present our findings in detail. The reporting is structured following the categories and subcategories of advice we identified through the affinity diagramming of our codes (see Table \ref{tab:taxonomy}). 
For each category, we analyze advice for individuals as well as recommendations made to companies and organizations. 
Figure \ref{fig:heatmap} shows an overview of the counts of assigned codes.
In cases where the advice for both target groups was very similar, we merge the reporting to avoid repetition. 
In addition, we provide noteworthy insights on advice sources where appropriate.

\begin{table}[tb]
\centering
\begin{threeparttable}
%\small
\setlength{\tabcolsep}{4pt}
\begin{tabularx}{\columnwidth}{lXrrr}
    \toprule
    \textbf{Target}&&Individuals & Organizations & Total\\
    \midrule
    \multicolumn{2}{l}{\textbf{Source}} &\var{codeCountBoth.T1} &\var{codeCountBoth.T2} & 372\\
    & Company & 42 & 26 & 57 \\
    & \acrshort{npo} & 36 & 8 & 44 \\
    & Government & 19 & 47 & 57\\
    & News & 44 & 19 & 53\\
    & Individual & 91 & 41 & 116\\
    \bottomrule
\end{tabularx}
\end{threeparttable}%
\caption{Overview of sources and targets of advice.}
\label{table:resource_source_target}
\end{table}
\subsubsection{Messaging \& Social Media}

The largest portion of advice (101 resources) targeted individuals and dealt with their social life online. 
We identified three key areas of advice on this topic: recommendations regarding secure instant messaging, advice on social media profiles and sharing practices, and pointers regarding misinformation. While some of the resources also addressed organizations, none of them directed this type of advice towards them.

\boldparagraph{Secure Messaging}

Recommendations regarding (secure) instant messaging were focused around which applications one should or should not use, with a total of 36 resources advocating for or against the use of at least one specific application. They mostly originated from \glspl{npo}, news outlets, and individuals and mentioned 13 distinct applications. Six resources warned that phone and SMS services were insecure and not private. Signal (9 times) and WhatsApp (7 times) were generally endorsed as secure, but there were also claims of insecurities that both companies called out as false. One individual tweeted that \blockquote[T2745]{@WhatsApp seems to be monitored by Russians,} to which WhatsApp said in their Twitter thread on the Ukraine war that \blockquote[T4048]{As always, your personal messages and calls are protected with end-to-end encryption by default so they cannot be intercepted by any government.} Similarly, there were claims that \blockquote[T2766]{Signal Russia has been breached.} 
Signal promptly refuted this:
\blockquote[T2763]{This is false. Signal is not hacked. We believe these rumors are part of a coordinated misinformation campaign meant to encourage people to use less secure alternatives.}

Telegram was the most discussed application (15 resources). All advice related to Telegram mentioned risks associated with the default settings of the application, which do not enable encryption of messages. Several also pointed out prevalent user misconceptions regarding this setting, \eg{}, that through \blockquote[T2757]{misleading marketing and press, most people [in Ukraine] believe it’s an encrypted app,} with one person taking it one step further claiming \blockquote[T2745]{that branding may literally cost lives.} Outside of highlighting the risks of using Telegram, a few news resources discussed the importance of the app in the distribution of information in Ukraine, both from individuals and government channels, stating that the uses may outweigh security concerns. Unfortunately, only one document provided a step-by-step guide to turning on encryption for chats in Telegram.

In general, the most frequently (11 times) recommended feature for secure instant messaging was (end-to end) encryption, followed by self-destructing messages (7 times), for which some companies specifically posted guides on how to turn them on. Peer-to-peer messaging applications were promoted as a means of communications in case of internet shut-downs or outages (7 times).

\boldparagraph{Advice for Social Media}

The advice around social media profiles and sharing practices centered on privacy and controlling what information people shared to whom. Individuals and social media companies were the main sources for this type of advice; the latter predominantly shared feature descriptions and usage guides for their own products. 
14 resources recommended that people should review their privacy settings or tighten visibility on their content. 
To this end, Meta introduced a region-bound new feature for locking Facebook profiles and hiding their content from the public that was recommended seven times. 
For Twitter users, deactivating their profile to hide old content was suggested four times. 
The measures were recommended to anyone in contact with people \blockquote[T4053]{in Ukraine to help protect people from being targeted,} and revealed a general concern that private information already available online may now lead to physical harm.

Advice on sharing practices called for being aware of what is shared (5 times). It extended from cautioning against posting sensitive information (6 times) to war-related specifics, \eg{}, location information and its potentially damaging role to military strategies was a focus. \blockquote[T8925]{Everyone is a target. DO NOT share locations of military operation in \#Ukraine in real time.} Accordingly, people were asked not to add meta data to posts (7 times), to remove meta data from previous posts (4 times), and not to live-tweet (4 times). One news article as well as one individual warned people against sharing videos or pictures of prisoners of war, \blockquote[D112]{which some experts have argued violates the Geneva Conventions.}

\boldparagraph{Bewaring of Misinformation}

Related to war news and information shared online, misinformation was a common topic. A total of 44 resources warned that wrong information was frequently shared and must be watched out for, with 16 specifically mentioning Russian disinformation. While most resources left it at this rather generic warning, some questions for spotting fake claims, such as \blockquote[T1004]{Does it look like Ukraine? Does it look like February time?} could be found, along with the advice to not share anything one had not verified (4 times). Two resources recommend reverse image searches to quickly find out if material had been put online previously in other contexts, and two resources recommended reporting accounts that shared fake information to combat its spread.

\begin{summaryBox}{Key Insights: Messaging \& Social Media}
    \begin{itemize}
        \item Secure messaging advice focused on the usage of specific applications.
        \item Social media advice focused on features to protect private information.
        \item Warnings about misinformation were common but often generic.
    \end{itemize}
\end{summaryBox}

\subsubsection{Safe Online Behavior}
At 84 resources, a significant portion of advice was on safe online behaviors and being careful with trust online. We divided this advice into the three major subcategories phishing, malware, and connections and anonymity. Most of this advice (71 resources) targeted individuals, while 26 resources addressed organizations and companies.

\boldparagraph{Phishing}
Phishing was widely considered to be a significant threat that would become more prevalent as scammers tried to profit from the war, with 54 resources calling for heightened vigilance of people (42 times) and organizations (23 times). The advice for both target groups was very similar, with companies being additionally told to spread the advice to their employees. Most advice came from government institutions (13 resources), companies (11 resources), and news outlets (11 resources).

The most general pieces of advice included to think before clicking (7 times), to not click links from unknown sources (3 times), to watch out for phishing (8 times), and to be suspicious of, \eg{}, unknown people, popups, requests, and things that are too good to be true (7 times). Seven resources advised to report any phishing attempts to authorities, and six resources cautioned against revealing personal information unless one was certain who was receiving them.

Many resources (35) regarded emails as the most likely medium for phishing. Of these, 19 generally said to be alert to phishing emails. More specifically, 10 resources noted that one should be suspicious of emails asking one to click links, with three resources going as far as saying one should not click links in emails at all. The sender was another aspect that advice focused on, recommending to be wary of false or unusual emails from trusted institutions (7 times), to verify suspicious senders and not trust sender addresses (2 times), and to not open emails from unknown senders (1 time).

Aside from email, resources also warned about phishing through instant messages (6 times) and social media platforms (3 times).

\boldparagraph{Malware}

We found that a rise in the threat of malicious software was widely reported as a consequence of the war (47 resources). Both individuals (23 times) and organizations (32 times) were warned about this threat in very similar ways, and a majority of the warnings originated from news articles (18 resources) and government institutions (9 resources), with the latter mainly targeting organizations and companies.

Of the 47 resources, 14 only generally talked about malware as a risk to be aware of, without providing mitigation strategies. The others focused on two main vectors through which malware could be introduced to a system: installing software, and email attachments.
Regarding email attachments, 12 resources advised varying degrees of cautiousness, ranging from being suspicious of email attachments (11 times) over not opening attachments from unknown senders (5 times) to not opening any unnecessary attachments (1 time).
For installing software (21 resources), the general advice of only installing software from trusted sources (11 resources) was extended in multiple ways specific to the crisis at hand. About half of the resources (11) discouraged the usage of software that came from Russia. A prominent example that most of them (9 resources) referenced was security software from the Russian provider Kaspersky, which multiple Western government agencies spoke out against, recommending \blockquote[D42]{replacing applications from Kaspersky's portfolio of antivirus software with alternative products over doubts about the reliability of the manufacturer}. Two resources asked people to beware of offers providing free software, like VPN services, pointing out that scammers may exploit people's acute need for such services to plant malicious software. One news article described how scammers also exploited people's wishes to help Ukraine by \blockquote[D9]{promot[ing] a fake DDoS tool on Telegram that installs a password and information-stealing trojan.}. Both this article and one Twitter user generally discouraged people from participating in any cyber attacks, as they are illegal and can be a big risk, especially to non-experts.

\boldparagraph{Connections \& Anonymity}

Advice regarding the safety of internet connections and being anonymous on the network appeared in 35 resources, of which 34 targeted individuals. Only six of the resources directed advice at organizations, with no remarkable differences in the advice for individuals. 

A majority (27 resources) recommended using specific types of software to secure connections and preserve privacy. The most common were VPN services (17 resources). Six of these were advertisements from a company providing VPN services. The others originated from \glspl{npo}, news outlets, and individuals. They described two different use cases of VPNs. One was to circumvent local censorship, telling people to \blockquote[D129]{set up VPN services to help you access blocked sites during a partial [internet] shutdown.} One person explained how they used \blockquote[T2893]{a VPN to a Western State to avoid Russian censorship.} The other was to secure communications and preserve anonymity, explaining that \blockquote[D140]{When configured correctly, a VPN will secure all of your communications
from local interception,} and \blockquote[D69]{It hides your IP address and your location. It also encrypts your data after leaving your device and traveling to whatever website you're visiting.}
Another software that \glspl{npo}, news outlets, and individuals commonly recommended for online anonymity was the TOR browser (12 resources).
It was seen as a tool to circumvent censorship, with one user tweeting \blockquote[T6901]{Tor is a mean of accessing truth safely. Tor is the equivalent of hidden atenas [sic] in the WWII.} Six tweets drew attention to a special project offering an uncensored and privacy protected way to browse Twitter using Tor.

Next to the software recommendations, there was also some general advice reminding people that internet connections could be a security risk (10 resources). Specifically, they said to not trust open networks (2 times) and to only use trusted networks (2 times). Additionally, it was recommended to turn off network features including WiFi, mobile internet, and Bluetooth whenever they were not used (3 times), as they may still disclose one's location. A very situation-specific advice that appeared twice was to hide Star Link ground stations Ukrainians received to ensure internet access and to use them sparingly, as they might become targets for military attacks.

\begin{summaryBox}{Key Insights: Safe Online Behavior}
    \begin{itemize}
        \item There were many warnings about intensified phishing and malware distribution but few actionable imperatives.
        \item To preserve confidentiality and anonymity, VPNs and Tor Browser were common suggestions.
    \end{itemize}
\end{summaryBox}

\subsubsection{Authentication}

In 59 resources talking about authentication, we identified three subcategories of security advice: advice regarding account credentials, recovery, and \gls{mfa}. Individuals (39 times) and organizations (31 times) were targeted by this advice alike, mostly by government agencies (17 times), companies (10 times), and individuals (10 times).

\boldparagraph{Passwords}
Security advice on credentials was mostly focused on passwords, with \var{codeCountBoth.Au1.2} resources advocating the use of strong passwords.
Specific criteria as to what constitutes a strong password were rarely given.
The resources mentioned randomness, length, and including a combination of letters, numbers, and special characters.
Using unique passwords for each account was mentioned a total of \var{codeCountBoth.Au1.1} times across our dataset, while fewer resources (\var{codeCountBoth.Au1.4}) recommended the use of a password manager. 
Of those recommending a password manager, eight emphasized that choosing strong passwords was still vital: \blockquote[D124]{Have a strong, unique password that you store in a password manager.}

The documents targeted at companies and organizations included policy directives such as ensuring that password policies are in place and being enforced: \blockquote[D20]{Ensure you have a strong password policy that is policed and maintained.}
Updating or resetting employees' passwords frequently was among the least frequently shared pieces of advice, with  \var{codeCountBoth.Au1.6.text} documents mentioning it.
In both cases, the advice was motivated in the context of possibly exposed credentials through security breaches via known vulnerabilities.

\boldparagraph{Recovery}

One company shared advice to their users on how to make account recovery more resilient to attacks by enabling a setting which requires either the associated email address or phone number to be entered on password reset attempts, whereas usually the username suffices.

In an attempt to thwart brute-force attacks and in response to a recent Russian state-sponsored attack on an \gls{ngo}, a government institution recommended the use of time-outs for repeated failed login attempts (D99).

\boldparagraph{Multi-Factor Authentication}
Advice around enabling \gls{mfa} was prevalent with \var{codeCountBoth.Au2} occurrences across all resources.
While the authors mentioned hardware tokens a few times, most chose to recommend \gls{2fa} without going into further detail.
Companies themselves recommended that their users enable \gls{mfa} for their services nine times, notable examples being Twitter and WhatsApp (Meta), while government institutions recommended it in a more general sense, mostly to businesses (14 times).
A notable exception is the following tweet that is targeted at the broader population:
\blockquote[T1183]{Implementing multi-factor authentication on your accounts makes it 99\% less likely you'll get hacked.}
In this case, the exaggerated claim of effectiveness might be an attempt to increase adoption, although it is supported by data from Microsoft \cite{weinert-2019:mfaefficacy}.

We identified 16 resources that mentioned enforcing \gls{mfa} for (privileged) accounts as a security measure for companies, while 20 mentioned it in total.
A majority of these resources (10) was documents shared by government institutions.
One individual on Twitter suggested an initial ``crash deployment'' of \gls{mfa} that would begin a staggered roll-out for privileged accounts and then extend it to accounts with access to confidential data as well as to employees who may be targeted by phishing attacks (T564).

\begin{summaryBox}{Key Insights: Authentication}
    \begin{itemize}
        % \item The most frequently recommended attributes for passwords were \blockquote{strong} and \blockquote{unique}.
        \item Choosing strong passwords was the most common advice but mostly lacked specific criteria.
        \item MFA was ascribed a very high effectiveness in securing accounts.
    \end{itemize}
\end{summaryBox}

\subsubsection{Hardware \& Software}

Advice regarding device security, mainly related to hardware and software, was split into three categories, namely: updating software and systems, using security software, and bolstering device and hardware security. The three categories of advice we found are discussed below. 

\boldparagraph{Software \& System Updates}

Updating software and devices regularly were some of the common pieces of advice that we encountered. Advice regarding updates for individuals included keeping general software and devices up-to-date (17 resources). Additionally, individuals were advised to install updates for friends and families (1 resource). \blockquote[T4098]{Update your own software on your phones, laptops, desktops, smart devices. Updates patch known security flaws. Once you've updated your own software, do it for your parents, aunts, uncles. This is actual self-defense.} Keeping systems and software updated and patched, and updating devices and device firmware was mentioned for organizations as well (10 resources). For instance, some of the advice indicated a sense of urgency in their messaging: \blockquote[T4064]{I cannot emphasize enough.  Everyone, all your companies, all your phones, everything, update your virus protection and download your security patches IMMEDIATELY.} Additionally, using automatic updates for devices and software was mentioned in six resources, and considering the availability of automatic updates when buying a new device was mentioned two times. A majority of the advice was also about reminding individuals and organizations to keep their security software like antivirus up-to-date (19 resources). 

\boldparagraph{Security Software}

In order to bolster the security of companies and individuals, three resources focused on advice to developers. 
This advice included building security into products from the \textit{ground-up}, conducting regular penetration tests and audits, and minimizing the complexity of systems and services used. \blockquote[D116]{Build security into your products from the ground up — ``bake it in, don’t bolt it on'' — to protect both your intellectual property and your customers’ privacy.}. 
In addition to advice provided to organizations, eight resources mentioned changing settings of anti-virus and anti-malware software to run periodic scans, checking for anti-virus signatures and patches, and disabling Microsoft macros (two resources each).
\blockquote[D31]{Implement mitigations against phishing and spear phishing attacks. Disable Microsoft Office macros by default and limit user privileges. Ensure that staff report all suspicious emails received, links clicked, or documents opened.} 
Additionally, a majority of the advice (21 resources) was also directed at individuals installing and using security software. 
Two resources mentioned \textit{use security software} without adding any further details.
However, using anti-virus software (10 times), anti-malware software (5 times), firewalls (8 times), vulnerability scanning software (3 times), spam-filtering services (3 times), integrity monitoring software (1 time), and a virtual machines (1 time) were specifically mentioned in the context of security software to strengthen individuals' security.

\boldparagraph{Device \& Hardware Security}

In terms of device and hardware security, some of the advice for individuals mentioned locking their devices. 
Locking devices without any further mention of the best practices or methods surfaced three times. Locking smartphones using a passcode or touch ID appeared two times. Furthermore, individuals were advised to set up auto-lock timers for their smartphones (1 resource). While some advice focused on using passcodes and touch IDs, we also found one resource catered to journalists that talked about disabling biometrics on devices: \blockquote[T972]{I can tell you that independent Russian newsrooms all instruct their employees in Russia to disable all biometrics on their smart devices, to prevent the cops from smashing your finger on Touch ID or holding your phone in front of you for Face ID.}  
Enabling biometrics for software access appeared once. 
Additionally, a facet of advice about being careful while plugging external devices into computers appeared in 10 resources. 
This advice further included scanning external devices before plugging them into personal computers (1 resource). 
Two resources also mentioned avoiding plugging external devices into computer systems altogether. 
Finally, one resource also mentioned using a data blocker before using USB charging ports. Other advice targeted towards individuals included discarding devices with security weaknesses (2 resources), factory resetting devices to remove malware (2 resources), factory resetting cellphones (1 resource), and keeping work and personal devices separate (2 resources). 
We encountered one resource from a news agency that provided advice for protecting devices in case people are arrested or detained and in case their devices may be \blockquote[D133]{confiscated and searched.} 
They specifically mentioned not using devices that are no longer supported by the manufacturer, using devices that support setting passwords, and enabling remote wiping of devices. 

Some of the resources (10) also advised turning off location services on devices to disable tracking. 
Additionally, one resource from an individual advised people to not use cell phones in safe houses.

\begin{summaryBox}{Key Insights: Hardware \& Software}
    \begin{itemize}
        \item Software advice focused on using security software such as antivirus and on making regular software secure from the beginning or through updates.
        \item Device security was centered around preventing unwanted access and tracking.
    \end{itemize}
\end{summaryBox}

\subsubsection{Storing Data}
Advice regarding what data to store and how appeared in 44 resources. 
It targeted both individuals (24 resources) and organizations (30 resources). 
We divided the advice into three distinct categories: backups, logging, and preventing unwanted access.

\boldparagraph{Backups}

A total of 30 resources advised individuals (16 resources) and organizations (23 resources) about backing up their data. 
Most prevalent was the general recommendation to have backups (22 resources), which was similar for both target groups. 
Contrastingly, the more specific advice differed. 
Organizations received recommendations tailored to professional data handling, such as testing the backup and restore process (12 resources) and isolating backups from their network (6 resources), \eg{} \blockquote[D89]{Test backup procedures to ensure that critical data can be rapidly restored if the organization is impacted
by ransomware or a destructive cyberattack; ensure that backups are isolated from network connections.}
Advice for individuals was more diverse and often focused on specific actions rather than a broader strategy. 
Examples include \blockquote[T534]{Scan or take photos of all important docs and send them to your own email account.} and \blockquote[D43]{Back up all your devices to an external hard drive or to the cloud.}

\boldparagraph{Logging}

Advice to store logging data was mainly given to organizations and companies (14 resources) and mostly originated from government institutions (11 times). 
The general advice to verify that logging was done and how and for how long the logs were stored (7 resources) was supplemented with pointers on what to log. 
Recommendations included key functions (4 resources), network activity (4 resources), authentication activity (2 resources), access to personnel information (1 resource), and changes to security-enabled groups (1 resource).

\boldparagraph{Preventing Access}

Of the nine resources giving advice on how to prevent unwanted access to data, all targeted individuals, and one also targeted organizations. 
A majority of them (6) came from \glspl{npo}. 
There were two strategies, minimizing how much data was stored and thus available (7 resources) and encrypting the data that was stored (7 resources). 
For data minimization, recommendations were to remove sensitive files from devices, not to store data unnecessarily, and to disconnect from accounts (4 resources each). 
Less common were pointers to rename contacts to hide their identity (2 resources) and to regularly delete one's browser history (1 resource). 
Advice on encryption focused on what to encrypt and named device data (7 resources), backups (4 resources), hard drives (2 resources), and cloud data (2 resources). 
One resource recommended to \blockquote[T3033]{activate the protocol to delete the information after a few wrong [decryption attempts].}

\begin{summaryBox}{Key Insights: Storing data}
    \begin{itemize}
        \item Advice focused on preserving data and preventing unwanted access.
        \item There was a stark contrast between professional strategies recommended to organizations and singular quick-and-easy actions recommended to individuals.
    \end{itemize}
\end{summaryBox}

\subsubsection{Organizational Policies}\label{subsubsec:company_advice}

We found 25 resources giving advice about policies that only applied to organizations and grouped them into the two categories incident response and recovery plans as well as access and network policies. About two thirds (15) of the resources gave advice coming from government organizations.

\boldparagraph{Incident Response \& Recovery Plans}

A vast majority of resources that dealt with organizational policies (23) contained information about responding to security incidents and having plans for recovery from such incidents. 
Of these, 11 recommended developing an incident response plan, while nine advised to verify that a plan existed and was up to date. 
Regarding the content of the plan, six resources said it should be known and actionable, and six stressed the importance of having contact information for essential personnel available. 
Three resources mentioned that routes of an incident response plan should be accessible even if systems had been shut down. 
Another three resources suggested practicing the response plan in the organization, \ie{}, the US agency CISA recommended to \blockquote[D89]{Conduct a tabletop exercise to ensure that all participants understand their roles during an incident.}

\boldparagraph{Access \& Network Policies}

A majority of resources (16) made recommendations regarding access control and network policies. For internal access, eight resources advised that the principle of least privilege access should be followed. Keeping track of authorization and timely removing accounts of leavers and unused accounts was recommended by eight resources as well. Two resources mentioned that next to digital access control, preventing unauthorized physical access to systems was an important measure. For networks, six resources advised to isolate them, and four resources recommended to disable any unused ports and protocols on the network. 
In addition to general network security measures, the US agency CISA included isolation and extra careful inspection of traffic from Ukrainian organizations and blocking activity from VPN or Tor connections into their situation-specific recommendations.

\begin{summaryBox}{Key Insights: Organizational Policies}
    \begin{itemize}
        \item Advice focused on up-to-date, properly communicated plans for incident response and recovery.
        \item Isolating networks and strict authorization were recommended defenses.
    \end{itemize}
\end{summaryBox}

\subsubsection{Learning \& Teaching}

Finally, advice on usage and distribution of security information, learning, and teaching was prevalent in our data collection, as various entities offered, referenced, and commented on advice resources. We collected 92 resources with such content, targeted at individuals (50 times) and organizations and companies (64 times). They had diverse sources, the most common being government agencies (24 resources) and individuals (21 resources), followed by \glspl{npo} (14 resources) and companies (13 resources). We identified four subcategories: meta advice about sharing security advice during crises, awareness and resources, learning, and building a threat model.

\boldparagraph{Recommendations for Sharing Advice During Crises}

Some of the tweets dealt with the topic of sharing security advice during crises itself, wherein the authors gave other professionals, who may want to share advice, guidance on how to prioritize classes of advice and what topics or phrasings to avoid.
% L1.3
Four authors of advice resources asked that the readers pass on the given advice to friends and family.
Advice givers should do their due diligence and refrain from recommending single tools while drastically overstating their efficacy with respect to security or privacy, especially during the current situation in Ukraine.
One Twitter user pointed out that giving digital security advice was a major responsibility and that \blockquote[T504]{[one should not] encourage people to entrust their safety to one thing. Especially not in conflict.}.

In line with this, \var{codeCountBoth.L3.2.text} resources encouraged others to give realistic as well as actionable advice that takes into account that security and privacy priorities may be different for people in Ukraine and that is more specific than, \eg{}, following all the advice that has been reiterated for years.
Correspondingly, one individual focused on actionable advice and called on companies to prioritize a fast roll-out of basic security measures in the face of emerging cyber threats: \blockquote[D112]{We need to make things BETTER, NOW! We can tweak and harden later, when we have the basics deployed.}.
Nine resources, which were mostly shared by companies citing government institutions (3) or by government institutions themselves (2), also recommended that companies raise awareness for increased risks by, \eg{}, performing employee training.
However, the resources mainly pointed to conveying current security best practices without going into further detail.

\boldparagraph{Awareness \& Resources}
58 resources did not offer advice themselves but rather raised awareness for resources provided by others.
\var{codeCountBoth.G8} resources offered help in the form of technical guidance or support, often directly to Ukrainian companies.
Government institutions were most notable here (14 times), followed by fellow companies and \glspl{npo} (8 times each).
For the former, this took the form of, \eg{}, accepting forwarded websites, emails, and texts in an effort to support Ukraine by not falling victim to attacks (T8889).
Companies offered free services like firewalls and VPNs. 
Offers for individual consulting on security were common as well.
In \var{codeCountBoth.G3.4.text} resources, the advice givers warned that \blockquote[T504]{there'll be well-intentioned twitter connectivity advice. Some great. Some not.} Others reported advice they had come across that may be impractical or even actively damaging to the individual's or company's security: \blockquote[T617]{[\ldots] Lots of great info but please don’t follow their mitigation advice for ICS. It’s not practical \& in some cases dangerous. [\ldots]}

Five resources in total advocated that companies and organizations follow current best practices in security without giving specifics.
Also in five cases, government institutions set up newsletters for companies to receive updates on emerging threats and advisories.
In contrast to the efforts around offering support, two Twitter users told companies and organizations that the steps to protect from cybercrime had not changed: \blockquote[T3978]{Contrary to the marketing emails that’ll flood your inbox in the coming days inviting you to a webcast on how to protect against Russian attacks, the measures to protect your org haven’t changed a bit since the war started.}

\boldparagraph{Learning}

General advice related to learning about security appeared in nine resources in total.

Staying up-to-date with security and privacy developments and to keep learning was shared in \var{codeCountBoth.L2.1.text} documents from companies, \glspl{npo}, government institutions, and news sites (1 time each).
D139 as a security guide for journalists is an example of a learning resource that became highly applicable again in light of the invasion.
It dedicates an entire section to technology security in conflict areas, ranging from threat modeling, secure communications, over mobile device security to malware, data integrity, and secure credentials.
Written by a \gls{npo} with a target group of journalists in general, it was shared again on Twitter by the \gls{npo}, this time specifically mentioning reporters in Ukraine.

Five documents, with four being targeted at companies, endorsed learning and getting advice from security experts as well as professionals.
In two, government institutions pointed to their own services, while a news site indicated urgency but stayed vague: \blockquote[D17]{If you don’t have a competent security team to help (and most don’t), you absolutely must find a reputable security partner immediately.}.

\boldparagraph{Threat Modeling}
Advice on building threat models as a foundation for choosing security advice to apply appeared in 28 resources and was a category with notable distinctions between advice targeted at individuals and advice for organizations.

Of the advice targeting individuals (14 resources), the majority (7 resources) generally recommended thinking about threats when making security choices and came from individuals (4 times) and \glspl{npo} (3 times). More specific pointers such as that average people may be targeted by advanced persistent threats or scammers and bots were rare (3 and 2 times) and came mostly from news outlets and government institutions. One individual stressed: \blockquote[T4098]{To a human scammer or a bot, they/it don't care who you are, you're just a vulnerable victim. Practice safe computing.}

Organizations and companies were targeted in 23 resources. For them, general pointers to think about threats were less prevalent (6 resources). Instead, most of the resources were more specific, with 10 referring to advanced persistent threats, three to the software supply chain as a potential attack vector, two to scammers and bots, and one to overseas attackers. Most of this information originated from government institutions (7 times) and was shared by news (4 times), individuals, \glspl{npo}, and the government itself (1 time each).

\begin{summaryBox}{Key Insights: Learning \& Teaching}
    \begin{itemize}
        \item Several resources called for giving advice responsibly and making it actionable.
        \item Offers for free individual support and consulting were extended to affected people and organizations.
        \item There was a disagreement between people calling for immediate measures and people saying the measures to take had not changed at all.
        \item Having a threat model was sometimes recommended, but there was no actionable guidance on prioritizing advice.
    \end{itemize}
\end{summaryBox}

\subsection{Comparison with Prior Work} \label{subsec:comparison}
To answer our second research question, we compare our findings for advice targeting individuals to those of two other papers that have investigated this kind of advice sharing. 
While \citeauthor{Reeder:2017:Steps} collected advice by asking experts to name the top three pieces of advice they would give non-tech-savvy users in 2017~\cite{Reeder:2017:Steps}, \citeauthor{Boyd:2021:Understanding} investigated advice shared on Twitter in the context of the \gls{blm} protests in 2020~\cite{Boyd:2021:Understanding}.
Additionally, we evaluate the advice from our data collection t hat was targeted at individuals using data from \citeauthor{Redmiles:2020:Comprehensive} on advice priority as well as uselessness and harmfulness of advice~\cite{Redmiles:2020:Comprehensive}.

\subsubsection{Comparing Data Collections}
In this section, we present the comparison of our data to that of prior work. The top ten most frequent pieces of advice from each data set can be found in Table \ref{table:toptenadvicecomparison}.

\boldparagraph{Advice for Non-tech-savvy Users}

In their analysis of advice for non-tech-savvy users. \citeauthor{Reeder:2017:Steps} collected and coded 231 expert responses for the advice they contained, using 152 unique codes. Of these, 56 match codes from our codebook. 
The frequencies with which advice appeared in their and our data collection were moderately correlated (Spearman, $r = 0.44$, $p = 0.004$), with notable similarities in the most frequent advice. Of their top ten, all pieces of advice were present in our data collection. Our top ten pieces of advice included their top four as well as one other of their pieces of advice. 
Of our top ten pieces of advice, those on the topic of misinformation, pointers to support with cyber security, insecurity of Telegram messenger, and VPN usage were not part of the data collected by \citeauthor{Reeder:2017:Steps}.

\boldparagraph{Advice Shared in the Context of BLM Protests}

Of the 193 unique codes \citeauthor{Boyd:2021:Understanding} assigned, only 26 matched codes in our codebook, which in part stems from them coding specifically for rationales of advice, while we did not. Of the matching advice, the frequencies in our data collection do not correlate with those in their collection in a statistically significant way (Spearman, $r = 0.278$, $p = 0.16$).
Of our top ten pieces of advice, only two occurred in their data. 
Of the advice that matched, all that belonged to the top ten during the \gls{blm} protests were present in our data collection, but only one was also among our top ten most frequent pieces of advice, with the others having low counts in our data collection.

\begin{summaryBox}{Key Insights: Comparing Data Collections}
    \begin{itemize}
        \item The most frequent advice was similar to that \citeauthor{Reeder:2017:Steps} found for non-tech-savvy users.
        \item There were few similarities between the advice around the \gls{event-short} and that shared during the \gls{blm} protests.
    \end{itemize}
\end{summaryBox}

\subsubsection{Evaluation of Advice}

In their paper ``A Comprehensive Quality Evaluation of Security and Privacy Advice on the Web'', \citeauthor{Redmiles:2020:Comprehensive} provide detailed insights into how end users and security experts evaluate security advice for end users from internet sources \cite{Redmiles:2020:Comprehensive}. 
We mapped 241 pieces of advice from their data set to \var{unique_codes} of our own unique codes. 
102 of their advice imperatives matched one of our \var{used_content_codes} assigned codes that refer to advice content. 
An overview of the top ten pieces of advice can be found in Table \ref{table:toptenadviceevaluation}.
%Excluding Target and Source codes

\boldparagraph{Priority Rankings}

With their replication package, \citeauthor{Redmiles:2020:Comprehensive} provide separate priority rankings of advice for expert and end user evaluation, which combine their more fine-grained ratings.
We look at the overall correlation between the rank a piece of advice got from experts as well as users and its frequency in our data. 
Additionally, we analyze the overlap and differences in the top ten advice pieces from each of the rankings and our data (see Table \ref{table:toptenadviceevaluation}).

We find that the expert priority ranking moderately correlates with the number of times advice was present in our data (Spearman, $r = 0.424$, $p < 0.0005$). Nine out of the top ten advice pieces from the expert ranking were present in our data collection. Two of them, namely \blockquote{Use different passwords for each account} and \blockquote{Use strong passwords}, were among our top ten most frequently shared advice.

The user priority ranking of advice only very weakly correlates with the frequencies in our data collection (Spearman, $r = 0.181$, $p = 0.004$). Of the top ten advice pieces, seven occurred in our data collection, albeit in overall much smaller quantities than those from the expert rankings. None could be found in our top ten most frequent pieces of advice.

In our top ten advice pieces, six were rated by users and experts. 
As expected given the correlations described before, they have rather high ranks in the expert ranking with the highest being first and the lowest at rank 53. 
By contrast, they are spread out through the user ranking with the highest at 22 and the lowest at 190. 
The other four top pieces of advice deal with misinformation, instant messenger recommendations, and pointers to sources of support. 
These topics are notably absent from the data \citeauthor{Redmiles:2020:Comprehensive} collected via user search queries and expert recommendations for advice sources in 2017.
The low correlation with the user ranking, which \citeauthor{Redmiles:2020:Comprehensive} based partly on user-perceived actionability of advice, aligns with our general finding that advice we collected was often very generic.

\boldparagraph{Useless or Harmful Advice}

Next to the priority rankings, \citeauthor{Redmiles:2020:Comprehensive} also provide data on which advice was deemed useless or harmful by at least one of the security experts they asked. 
Of the 25 pieces of advice labeled useless, only \blockquote{shut down your computer} was part of our data collection (1 resource). 
Of the advice categorized as harmful, four pieces can be found in our data: \blockquote{Use Tor} (12 resources), \blockquote{don't open attachments from unknown senders} (5 resources), \blockquote{change passwords frequently} (1 resource), and \blockquote{buy devices with passwords} (1 resource).
This shows that usefulness is a debated topic. 
To some extent, advice that some experts have deemed harmful or useless will continue to be circulated.

\begin{summaryBox}{Key Insights: Evaluation of Advice}
    \begin{itemize}
        \item Advice frequently shared during the \gls{event-short} was given high priority by experts in the study of \citeauthor{Redmiles:2020:Comprehensive}~\cite{Redmiles:2020:Comprehensive}.
        \item Much of the very frequent advice scored low on a user priority ranking based on i.a. perceived actionability and time consumption~\cite{Redmiles:2020:Comprehensive}.
    \end{itemize}
\end{summaryBox}

\begin{table*}[tbph!]
\centering
\begin{threeparttable}
\footnotesize
\rowcolors{2}{white}{gray!10}
\setlength{\tabcolsep}{3pt}
\begin{tabular}{rp{.3\textwidth}p{.3\textwidth}p{.3\textwidth}}
\toprule
\textbf{\#} & \textbf{Our Data} & \textbf{Reeder et al. \cite{Reeder:2017:Steps}} & \textbf{Boyd et al. \cite{Boyd:2021:Understanding}}\\
\midrule
1 & Beware of disinformation&Keep systems and software up to date&Disable biometric unlocking\\
2 & Support pointers&Use unique passwords &Use E2EE messaging app\\
3 & Use strong passwords&Use strong passwords &Use Signal\\
4 & Use multi-factor authentication&Use multi-factor authentication&Turn off location\\
5 & Keep systems and software up-to-date/patched&Use antivirus software&Avoid identifiable people (in social media posts)\\
6 & Use a VPN&Use a password manager&Remove metadata (from social media posts)\\
7 & Beware of Russian disinformation&Use HTTPS&Encrypt device\\
8 & Be alert to phishing email&Use only software from trusted sources&Turn off Bluetooth and WiFi\\
9 & Telegram is insecure&Use automatic updates&Use a strong password\\
10 & Use different passwords for each account&Be careful/think before you click&Disconnect cellular data\\
\bottomrule
\end{tabular}
\end{threeparttable}
\caption{Top ten pieces of advice compared by frequency of appearance.}
\label{table:toptenadvicecomparison}
\end{table*}

\begin{table*}[tbph!]
\centering
\begin{threeparttable}
\footnotesize
\rowcolors{2}{white}{gray!10}
\setlength{\tabcolsep}{3pt}
\begin{tabular}{rp{.3\textwidth}p{.3\textwidth}p{.3\textwidth}}
\toprule
\textbf{\#} & \textbf{Our Data} & \textbf{Expert Priority~\cite{Redmiles:2020:Comprehensive}} & \textbf{User Priority~\cite{Redmiles:2020:Comprehensive}}\\
\midrule
1 & Beware of disinformation&Use different passwords for each account&Never give your credentials to third parties\\
2 & Support pointers&Update devices and device firmware&Buy devices with security-focused platforms\\
3 & Use strong passwords&Use anti-malware software&Don't open unnecessary attachments\\
4 & Use multi-factor authentication&Scan attachments you open for viruses&Use anti-virus software\\
5 & Keep systems and software up-to-date/patched&Never give your credentials to third parties&Don't click random or unfamiliar links from unknown senders\\
6 & Use a VPN&Use unique passwords for different accounts&Verify suspicious emails, senders, and email contents\\
7 & Beware of Russian disinformation&Use (end-to-end) encryption for communication&Not open email from unknown senders\\
8 & Be alert to phishing email&Keep anti-virus software installed and up-to-date&Don't friend/put in your contacts people you don't know\\
9 & Telegram is insecure&Use strong passwords&Be suspicious if something is too good to be true\\
10 & Use different passwords for each account&Turn on automatic updates for devices&Set your antivirus/anti-malware to run periodic full scans\\
\bottomrule
\end{tabular}
\end{threeparttable}
\caption{Top ten pieces of advice compared to user and expert priority rankings in \citeauthor{Redmiles:2020:Comprehensive}~\cite{Redmiles:2020:Comprehensive}.}
\label{table:toptenadviceevaluation}
\end{table*}

%% file: sections/07-discussion.tex
We analyzed \var{tweets.coded} tweets and \var{resources.total} linked documents shared around the \gls{event} regarding the security and privacy advice they contained.
In \var{used_content_codes} unique pieces of advice, we find a large variety of recommendations that five different types of sources gave to individuals and organizations.
In addition, we find noteworthy similarities to the advice for non-tech-savvy users by \citeauthor{Reeder:2017:Steps}~\cite{Reeder:2017:Steps}, and significant differences to advice collected in the context of BLM protests by \citeauthor{Boyd:2021:Understanding}~\cite{Boyd:2021:Understanding}.
We note a stronger correlation of advice frequency to the priority ranking of experts than to that of users presented by \citeauthor{Redmiles:2020:Comprehensive}~\cite{Redmiles:2020:Comprehensive}.
From our findings, we derive the following main insights.

\boldparagraph{The lack of prioritization becomes even more apparent and detrimental during a crisis}
The \var{used_content_codes} unique pieces of advice in our data collection show that the security advice shared around the \gls{event-short} lacks focus. A plethora of companies, \glspl{npo}, government agencies, news outlets, and individuals shared advice they believed relevant, resulting in a large amount of advice without any obvious way to prioritize. 
The various sources failed to find a consensus or rally behind a common set of measures to take.
Frequently, they stressed the importance of following their advice immediately to counter the threat, contributing to an abundance of advice marked as high priority. 
Much of this advice is too generic to be actionable and a reiteration of long-established security measures that have been shown to struggle with adoption in the past \cite{Redmiles:2016:Learned}.
In this, the advice shows similarities to that collected by \citeauthor{Reeder:2017:Steps}~\cite{Reeder:2017:Steps} and \citeauthor{Redmiles:2016:Learned}~\cite{Redmiles:2016:Learned}. 
Both find a need for a clear, consistent set of advice to provide to end users \cite{Reeder:2017:Steps,Redmiles:2020:Comprehensive}. 
We find that advice for organizations could similarly profit from consistency and prioritization.

In the face of an acute crisis, it seems even more detrimental to overload advice recipients, as the additional pressure during a crisis can be expected to make it much harder to deal with. 
We argue that in a war zone and during other crises, the amount of time and resources that can be spent on the implementation of measures is even more constrained than usual for both individuals and organizations. 
Being overwhelmed is not helpful and may add additional pressure. 
We suggest the involvement of people with relevant experiences in the development of guidelines to giving advice during crises, to account for the special circumstances recipients of the advice might face. 

In addition, we re-emphasize the need for empirical research into the impact of security measures with the goal of establishing an agreed-upon core set of high priority security and privacy advice such as suggested by \citeauthor{Redmiles:2020:Comprehensive}~\cite{Redmiles:2020:Comprehensive}.
If tailored to each target group, such a core set could then serve as a baseline for advice given to both individuals and organizations.

\boldparagraph{Offers for individual support and counseling can be a valuable tool}
In addition to advice imperatives, we found offers for free security support and counseling on security threats and countermeasures. 
In contrast to broadcast guidelines, these can take individual needs and circumstances into account and can be a highly effective method if conducted by qualified experts. 
We suggest investigating the adoption and effectiveness of such offers in future work and to consider funding them as long-term initiatives if proven successful. 
In the meantime, we encourage the security community to keep making such generous offers to people facing war and other crises.

\boldparagraph{Misinformation is a rising threat}
While misinformation, and specifically misinformation on social media during crises, has been studied in various contexts \cite{Starbird:2015:Connected, Starbird:2019:Nature, Houlden:2021:Covid}, the topic was not present in any of the related work on security and privacy advice that we compared our data to.
Prior work finds a lack of actionable countermeasures \cite{Starbird:2019:Nature}, and we argue that this is an important issue that should concern the information security community. 
While the focus previously appeared to lie with political opinions and real life events, the claim that Signal had been breached and the subsequent dismissal of that statement as misinformation by the company show that security advice and practices can be influenced just as much.
Misinformation pointers were a prevalent topic in our analysis and are an important factor for safe social media usage. 
During a war or crisis, access to information can be vital, and the ability to tell information and misinformation apart is thus critical \cite{Starbird:2015:Connected,Starbird:Boston:2014}. Unfortunately, actionable advice on how to do that was rare in our resources. 
In addition, the given measures such as verifying details and reverse image searches are time-consuming and thus unlikely to be adopted for the many posts and many pieces of advice that users see~\cite{Geeng:Fakenews:2020}.
As a research community, we should contribute to combating misinformation by measuring its impact on security behaviors and developing mitigation strategies for security experts as well as users.

%% file: sections/08-conclusion.tex
In this paper, we studied security and privacy advice that was shared around the \gls{event}.
Specifically, we analyzed \var{tweets.coded} tweets and \var{resources.coded} linked documents using qualitative open coding. We distinguished advice targeted at individuals and organizations, as well as five types of sources: companies, \glspl{npo}, government agencies, news outlets, and individuals.
Using affinity diagramming, we created a taxonomy containing \var{used_content_codes} unique pieces of advice, clustered into seven categories.
We then compared our findings to those of three prior studies, finding significant similarities to two of them as well as newly emerging topics.
Unfortunately, we confirm previous findings that overwhelming amounts of advice are shared. 
Most are called high priority, leaving no pointers for effective prioritization. 
This appears even more detrimental in stressful and resource-constrained situations like wars and other crises. 
We find that in light of this, offers for individual support could be a valuable tool, the adoption, and effectiveness of which should be part of further research. 
In addition, we identify misinformation as a rising threat that may corrupt efforts to educate populations on security and privacy measures. 
As such, it should be addressed by the information security community in future research on its impact, and possible mitigation strategies and their effectiveness.

%% file: sections/09-appendix.tex
\section{Taxonomy}\label{sec:taxonomy}
The table below shows the categories and subcategories of our taxonomy, as well as the two most frequently assigned codes as examples. 
The column \emph{count} contains the total number of resources a code was assigned to. For the categories and subcategories, the count is an aggregate of all codes they contain. 
The columns \emph{Individuals} and \emph{Organizations} contain the number of resources that were assigned the respective code as well as this target. 

\begin{table*}[tbph!]
\centering
\begin{threeparttable}
\footnotesize
%\rowcolors{1}{white}{gray!10}
\setlength{\tabcolsep}{3pt}
\begin{tabular}{p{28mm}p{40mm}p{65mm}rrr}
\toprule
\textbf{Category}                & \textbf{Subcategory}                      & \textbf{Code}                                                                                                                                  & \textbf{Count} & \textbf{Individuals} & \textbf{Organizations}  \\
\midrule
\textbf{Messaging\,\&\,Social \nobreak Media} &                                  &                                                                                                                                       & 104   & 101         & 9              \\
                        & Secure Messaging &                                                                                                                                       & 44    & 43          & 3              \\
                        &                                  & Don't use Telegram/Telegram is insecure                                                                                             & 15    & 15          & 1              \\
                        &                                  & Use (end-to-end) encryption for communication                                                                                         & 11    & 11          & 1              \\
                        & Advice for Social Media          &                                                                                                                                       & 31    & 31          & 1              \\
                        &                                  & Review privacy settings                                                                                                               & 12    & 12          & 0              \\
                        &                                  & Don't post photos/test with metadata                                                                                                & 7     & 7           & 0              \\
                        & Misinformation                   &                                                                                                                                       & 44    & 42          & 6             \\
                        &                                  & Disinformation                                                                                                                        & 28    & 26          & 6              \\
                        &                                  & Beware of Russian disinformation                                                                                                      & 16    & 16          & 0              \\
\midrule
\textbf{Safe online behavior}   &                                  &                                                                                                                                       & 84    & 71          & 26             \\
                        & Phishing                         &                                                                                                                                       & 54    & 42          & 23             \\
                        &                                  & Be alert to phishing email                                                                                                           & 19    & 15          & 7              \\
                        &                                  & Be suspicious of emails asking you to click links                                                                                     & 10    & 9           & 4              \\
                        & Malware                          &                                                                                                                                       & 47    & 23          & 32             \\
                        &                                  & Don't use software from Russia                                                                                                        & 11    & 6           & 10             \\
                        &                                  & Be suspicious of attachments                                                                                                          & 11    & 10          & 5              \\
                        & Connections \& Anonymity           &                                                                                                                                       & 35    & 34          & 6              \\
                        &                                  & Use a VPN                                                                                                                             & 17    & 16          & 3              \\
                        &                                  & Use anonymity systems (Use TOR/Psiphon)                                                                                               & 12    & 12          & 2              \\
\midrule
\textbf{Authentication}          &                                  &                                                                                                                                       & 59    & 39          & 31             \\
                        & Passwords                        &                                                                                                                                       & 40    & 29          & 17             \\
                        &                                  & Use strong passwords                                                                                                                  & 34    & 24          & 15             \\
                        &                                  & Use different passwords for each account                                                                                              & 18    & 14          & 7              \\
                        & Recovery                         &                                                                                                                                       & 4     & 3         & 1            \\
                        &                                  & Require email and phone number for a password reset                                                                                   & 3     & 3           & 0              \\
                        &                                  & Enable timeouts and lock-outs for failed log-in attempts                                                                              & 1     & 0           & 1              \\
                        & Multi-Factor Authentication      &                                                                                                                                       & 45    & 28          & 26             \\
                        &                                  & Use MFA                                                                                                                               & 30    & 20          & 15             \\
                        &                                  & Enforce MFA for privileged accounts\slash services\slash systems                                                                    & 20    & 9           & 16             \\
\midrule
\textbf{Hardware \& Software}      &                                  &                                                                                                                                       & 66    & 42          & 37             \\
                        & Software \& System Updates         &                                                                                                                                       & 42    & 22          & 31             \\
                        &                                  & Keep systems/software up to date                                                                                                      & 34    & 17          & 27             \\
                        &                                  & Keep anti-virus software installed and up-to-date                                                                                     & 11    & 7           & 6              \\
                        & Security Software                &                                                                                                                                       & 27    & 12          & 21             \\
                        &                                  & Use anti-virus software                                                                                                               & 10    & 7           & 7              \\
                        &                                  & Use anti-malware software                                                                                                             & 5     & 3           & 4              \\
                        & Device and Hardware Security     &                                                                                                                                       & 22    & 19          & 4              \\
                        &                                  & Turn off location devices                                                                                                             & 10    & 10          & 0              \\
                        &                                  & Lock devices                                                                                                                          & 3     & 3           & 0              \\
\midrule
\textbf{Storing Data}            &                                  &                                                                                                                                       & 44    & 24          & 30             \\
                        & Backups                          &                                                                                                                                       & 30    & 16          & 23             \\
                        &                                  & Backup your data                                                                                                                      & 22    & 13          & 18             \\
                        &                                  & Test backup/restore                                                                                                                 & 13    & 4           & 12             \\
                        & Logging                          &                                                                                                                                       & 17    & 4           & 14             \\
                        &                                  & Ensure logging is done, storage, retention periods                                                           & 7     & 0           & 7              \\
                        &                                  & Log key functions                                                                                                                     & 4     & 0           & 4              \\
                        & Preventing Access                &                                                                                                                                       & 9     & 9           & 1              \\
                        &                                  & Encrypt your device data                                                                                                              & 5     & 5           & 1              \\
                        &                                  & Log out of accounts                                                                                                                   & 4     & 4           & 0              \\
\midrule
\textbf{Organizational Policies} &                                  &                                                                                                                                       & 25    & 5           & 24             \\
                        & Incident \& Recovery Plans         &                                                                                                                                       & 23    & 5           & 22             \\
                        &                                  & Incident Response Plans                                                                                                               & 11    & 3           & 10             \\
                        &                                  & Verify an incident response plan exists and is up to date                                                                              & 9     & 1           & 9              \\
                        & Access \& Network Policies         &                                                                                                                                       & 16    & 0           & 16             \\
                        &                                  & Apply least privilege access                                                                                                          & 8     & 0           & 8              \\
                        &                                  & Track authorization and access, remove leavers                                                                                        & 7     & 0           & 7              \\
\midrule
\textbf{Learning \& Teaching}       &                                  &                                                                                                                                       & 92    & 50          & 64             \\
                        & Recommendations for Sharing Advice During Crises     &                                                                                                                                       & 18    & 11          & 13             \\
                        &                                  & Alert users about increased risks                                                                                                     & 9     & 3           & 9              \\
                        &                                  & Share advice with friends and family                                                                                                  & 4     & 3           & 1              \\
                        & Awareness \& Resources             &                                                                                                                                       & 58    & 31          & 40             \\
                        &                                  & Support pointers                                                                                                                      & 43    & 25          & 27             \\
                        &                                  & Guidelines                                                                                                                            & 5     & 1           & 5              \\
                        & Learning                         &                                                                                                                                       & 9     & 5           & 6              \\
                        &                                  & Seek professional help for cyber security issues                                                                                       & 5     & 2           & 4              \\
                        &                                  & Always keep learning about security and privacy                                                                                       & 4     & 3           & 2              \\
                        & Threat Modeling                  &                                                                                                                                       & 28    & 14          & 23             \\
                        &                                  & Threat model                                                                                                                          & 10    & 7           & 6              \\
                        &                                  & Advanced persistent threat groups                                                                                                     & 10    & 2           & 10             \\
\bottomrule
\end{tabular}
\end{threeparttable}
\caption{Taxonomy.}
\label{tab:taxonomy}
\end{table*}

%% file: references/refs.bib
@article{Makhortykh:2015:Savedonbasspeople,
    title       = {\#SaveDonbassPeople: Twitter, propaganda, and conflict in Eastern Ukraine},
    author      = {Makhortykh, Mykola and Lyebyedyev, Yehor},
    journal     = {The Communication Review},
    volume      = {18},
    number      = {4},
    pages       = {239--270},
    year        = {2015},
    publisher   = {Taylor \& Francis}
}

@inproceedings{Ronzhyn:2014:Use,
    title       = {The use of Facebook and Twitter during the 2013-2014 protests in Ukraine},
    author      = {Ronzhyn, Alexander},
    booktitle   = {Proceedings of the European Conference on Social Media},
    pages       = {442--449},
    year        = {2014}
}

@article{Pantti:2019:Personalisation,
    title       = {The personalisation of conflict reporting: Visual coverage of the Ukraine crisis on Twitter},
    author      = {Pantti, Mervi},
    journal     = {Digital Journalism},
    volume      = {7},
    number      = {1},
    pages       = {124--145},
    year        = {2019},
    publisher   = {Taylor \& Francis}
}

@inproceedings{Mishler:2015:Using,
    title           = {Using structural topic modeling to detect events and cluster Twitter users in the Ukrainian crisis},
    author          = {Mishler, Alan and Crabb, Erin Smith and Paletz, Susannah and Hefright, Brook and Golonka, Ewa},
    booktitle       = {International conference on human-computer interaction},
    pages           = {639--644},
    year            = {2015},
    organization    = {Springer}
}

@article{Suslov:2014:Crimea,
    title       = {{``Crimea Is Ours!''} Russian popular geopolitics in the new media age},
    author      = {Suslov, Mikhail D},
    journal     = {Eurasian geography and economics},
    volume      = {55},
    number      = {6},
    pages       = {588--609},
    year        = {2014},
    publisher   = {Taylor \& Francis}
}

@article{Wiggins:2016:Crimea,
    title       = {Crimea River: Directionality in memes from the Russia-Ukraine conflict},
    author      = {Wiggins, Bradley E},
    journal     = {International Journal of Communication},
    volume      = {10},
    pages       = {35},
    year        = {2016}
}

@inproceedings{Herley:2009:Advice,
  title     = {So long, and no thanks for the externalities: the rational rejection of security advice by users},
  author    = {Herley, Cormac},
  booktitle = {Proceedings of the 2009 workshop on New security paradigms workshop},
  pages     = {133--144},
  year      = {2009}
}

@inproceedings{Redmiles:2016:Learned,
    title       = {How i learned to be secure: a census-representative survey of security advice sources and behavior},
    author      = {Redmiles, Elissa M and Kross, Sean and Mazurek, Michelle L},
    booktitle   = {Proceedings of the 2016 ACM SIGSAC Conference on Computer and Communications Security},
    pages       = {666--677},
    year        = {2016}
}

@inproceedings{Das:2014:Effect,
    title       = {The effect of social influence on security sensitivity},
    author      = {Das, Sauvik and Kim, Tiffany Hyun-Jin and Dabbish, Laura A and Hong, Jason I},
    booktitle   = {10th Symposium On Usable Privacy and Security (SOUPS 2014)},
    pages       = {143--157},
    year        = {2014}
}

@inproceedings{Das:2014:Increasing,
    title       = {Increasing security sensitivity with social proof: A large-scale experimental confirmation},
    author      = {Das, Sauvik and Kramer, Adam DI and Dabbish, Laura A and Hong, Jason I},
    booktitle   = {Proceedings of the 2014 ACM SIGSAC conference on computer and communications security},
    pages       = {739--749},
    year        = {2014}
}

@inproceedings{Howe:2012:Psychology,
    title           = {The psychology of security for the home computer user},
    author          = {Howe, Adele E and Ray, Indrajit and Roberts, Mark and Urbanska, Malgorzata and Byrne, Zinta},
    booktitle       = {2012 IEEE Symposium on Security and Privacy},
    pages           = {209--223},
    year            = {2012},
    organization    = {IEEE}
}

@inproceedings{Ion:2015:no,
    title       = {{`... No one Can Hack My Mind'}: Comparing Expert and {Non-Expert} Security Practices},
    author      = {Ion, Iulia and Reeder, Rob and Consolvo, Sunny},
    booktitle   = {Eleventh Symposium On Usable Privacy and Security (SOUPS 2015)},
    pages       = {327--346},
    year        = {2015}
}

@inproceedings{Busse:2019:Replication,
    title       = {Replication: No One Can Hack My Mind Revisiting a Study on Expert and {Non-Expert} Security Practices and Advice},
    author      = {Busse, Karoline and Sch{\"a}fer, Julia and Smith, Matthew},
    booktitle   = {Fifteenth Symposium on Usable Privacy and Security (SOUPS 2019)},
    pages       = {117--136},
    year        = {2019}
}

@inproceedings{Redmiles:2016:Think,
    title           = {I think they're trying to tell me something: Advice sources and selection for digital security},
    author          = {Redmiles, Elissa M and Malone, Amelia R and Mazurek, Michelle L},
    booktitle       = {2016 IEEE Symposium on Security and Privacy (SP)},
    pages           = {272--288},
    year            = {2016},
    organization    = {IEEE}
}

@inproceedings{Fagan:2016:They,
    title       = {Why do they do what they do?: A study of what motivates users to (not) follow computer security advice},
    author      = {Fagan, Michael and Khan, Mohammad Maifi Hasan},
    booktitle   = {Twelfth symposium on usable privacy and security (SOUPS 2016)},
    pages       = {59--75},
    year        = {2016}
}

@inproceedings{Redmiles:2020:Comprehensive,
    title       = {A comprehensive quality evaluation of security and privacy advice on the web},
    author      = {Redmiles, Elissa M and Warford, Noel and Jayanti, Amritha and Koneru, Aravind and Kross, Sean and Morales, Miraida and Stevens, Rock and Mazurek, Michelle L},
    booktitle   = {29th USENIX Security Symposium (USENIX Security 20)},
    pages       = {89--108},
    year        = {2020}
}

@inproceedings{Acar:2017:Developers,
    title           = {Developers need support, too: A survey of security advice for software developers},
    author          = {Acar, Yasemin and Stransky, Christian and Wermke, Dominik and Weir, Charles and Mazurek, Michelle L and Fahl, Sascha},
    booktitle       = {2017 IEEE Cybersecurity Development (SecDev)},
    pages           = {22--26},
    year            = {2017},
    organization    = {IEEE}
}

@inproceedings{Boyd:2021:Understanding,
    title       = {Understanding the Security and Privacy Advice Given to Black Lives Matter Protesters},
    author      = {Boyd, Maia J and Sullivan Jr, Jamar L and Chetty, Marshini and Ur, Blase},
    booktitle   = {Proceedings of the 2021 CHI Conference on Human Factors in Computing Systems},
    pages       = {1--18},
    year        = {2021}
}

@article{Tahaei:2022:Understanding,
    title       = {Understanding Privacy-Related Advice on Stack Overflow},
    author      = {Tahaei, Mohammad and Li, Tianshi and Vaniea, Kami},
    journal     = {Proceedings on Privacy Enhancing Technologies},
    volume      = {1},
    pages       = {18},
    year        ={2022}
}

@inproceedings{Beris:2015:Employee,
    title       = {Employee rule breakers, excuse makers and security champions: mapping the risk perceptions and emotions that drive security behaviors},
    author      = {Beris, Odette and Beautement, Adam and Sasse, M Angela},
    booktitle   = {Proceedings of the 2015 New Security Paradigms Workshop},
    pages       = {73--84},
    year        = {2015}
}

@inproceedings{Das:2019:Typology,
    title       = {A Typology of Perceived Triggers for {End-User} Security and Privacy Behaviors},
    author      = {Das, Sauvik and Dabbish, Laura A and Hong, Jason I},
    booktitle   = {Fifteenth Symposium on Usable Privacy and Security (SOUPS 2019)},
    pages       = {97--115},
    year        = {2019}
}

@inproceedings{Nicholson:2019:Older,
    title       = {{``If It's Important It Will Be A Headline''}: Cybersecurity Information Seeking in Older Adults},
    author      = {Nicholson, James and Coventry, Lynne and Briggs, Pamela},
    booktitle   = {Proceedings of the 2019 CHI Conference on Human Factors in Computing Systems},
    pages       = {1--11},
    year        = {2019}
}

@article{Reeder:2017:Steps,
    title       = {152 simple steps to stay safe online: Security advice for non-tech-savvy users},
    author      = {Reeder, Robert W and Ion, Iulia and Consolvo, Sunny},
    journal     = {IEEE Security \& Privacy},
    volume      = {15},
    number      = {5},
    pages       = {55--64},
    year        = {2017},
    publisher   = {IEEE}
}

@article{Ermoshina:2017:Migrating,
    title       = {Migrating servers, elusive users: Reconfigurations of the Russian Internet in the post-Snowden era},
    author      = {Ermoshina, Ksenia and Musiani, Francesca},
    journal     = {Media and Communication},
    volume      = {5},
    number      = {1},
    pages       = {42--53},
    year        = {2017}
}

@article{Stanton:2016:Security,
    title       = {Security fatigue},
    author      = {Stanton, Brian and Theofanos, Mary F and Prettyman, Sandra Spickard and Furman, Susanne},
    journal     = {It Professional},
    volume      = {18},
    number      = {5},
    pages       = {26--32},
    year        = {2016},
    publisher   = {IEEE}
}

@inproceedings{Das:2015:Role,
    title       = {The role of social influence in security feature adoption},
    author      = {Das, Sauvik and Kramer, Adam D. and Dabbish, Laura A. and Hong, Jason I.},
    booktitle   = {Proceedings of the 18th ACM conference on computer supported cooperative work \& social computing},
    pages       = {1416--1426},
    year        = {2015}
}

@inproceedings{Redmiles:2018:Dancing,
    title       = {Dancing pigs or externalities? Measuring the rationality of security decisions},
    author      = {Redmiles, Elissa M and Mazurek, Michelle L and Dickerson, John P},
    booktitle   = {Proceedings of the 2018 ACM Conference on Economics and Computation},
    pages       = {215--232},
    year        = {2018}
}

@inproceedings{Burke:2009:Feed,
    title       = {Feed me: motivating newcomer contribution in social network sites},
    author      = {Burke, Moira and Marlow, Cameron and Lento, Thomas},
    booktitle   = {Proceedings of the SIGCHI conference on human factors in computing systems},
    pages       = {945--954},
    year        = {2009}
}

@article{Aytes:2003:Research,
    title       = {A research model for investigating human behavior related to computer security},
    author      = {Aytes, Kregg and Conolly, Terry},
    booktitle   = {Proceedings of the Ninth Americas Conference on Information Systems (AMCIS)},
    year        = {2003},
    month       = aug
}

@inproceedings{Egelman:2008:You,
    title       = {You've been warned: an empirical study of the effectiveness of web browser phishing warnings},
    author      = {Egelman, Serge and Cranor, Lorrie Faith and Hong, Jason},
    booktitle   = {Proceedings of the SIGCHI Conference on Human Factors in Computing Systems},
    pages       = {1065--1074},
    year        = {2008}
}

@article{Acquisti:2017:Nudges,
    title       = {Nudges for privacy and security: Understanding and assisting users’ choices online},
    author      = {Acquisti, Alessandro and Adjerid, Idris and Balebako, Rebecca and Brandimarte, Laura and Cranor, Lorrie Faith and Komanduri, Saranga and Leon, Pedro Giovanni and Sadeh, Norman and Schaub, Florian and Sleeper, Manya and others},
    journal     = {ACM Computing Surveys (CSUR)},
    volume      = {50},
    number      = {3},
    pages       = {1--41},
    year        = {2017},
    publisher   = {ACM New York, NY, USA}
}

@inproceedings{Bravo:2013:Your,
    title       = {Your attention please: Designing security-decision UIs to make genuine risks harder to ignore},
    author      = {Bravo-Lillo, Cristian and Komanduri, Saranga and Cranor, Lorrie Faith and Reeder, Robert W and Sleeper, Manya and Downs, Julie and Schechter, Stuart},
    booktitle   = {Proceedings of the Ninth Symposium on Usable Privacy and Security},
    pages       = {1--12},
    year        = {2013}
}

@inproceedings{Frik:2018:Better,
    title       = {Better late (r) than never: increasing cyber-security compliance by reducing present bias},
    author      = {Frik, Alisa and Egelman, Serge and Harbach, Marian and Malkin, Nathan and Peer, Eyal},
    booktitle   = {Symposium on Usable Privacy and Security},
    pages       = {12--14},
    year        = {2018}
}

@article{Rader:2015:Identifying,
    title       = {Identifying patterns in informal sources of security information},
    author      = {Rader, Emilee and Wash, Rick},
    journal     = {Journal of Cybersecurity},
    volume      = {1},
    number      = {1},
    pages       = {121--144},
    year        = {2015},
    publisher   = {Oxford University Press}
}

@inproceedings{Zou:2020:Examining,
    title       = {Examining the adoption and abandonment of security, privacy, and identity theft protection practices},
    author      = {Zou, Yixin and Roundy, Kevin and Tamersoy, Acar and Shintre, Saurabh and Roturier, Johann and Schaub, Florian},
    booktitle   = {Proceedings of the 2020 CHI Conference on Human Factors in Computing Systems},
    pages       = {1--15},
    year        = {2020}
}

@inproceedings{Dekoven:2019:Measuring,
    title       = {Measuring security practices and how they impact security},
    author      = {DeKoven, Louis F and Randall, Audrey and Mirian, Ariana and Akiwate, Gautam and Blume, Ansel and Saul, Lawrence K and Schulman, Aaron and Voelker, Geoffrey M and Savage, Stefan},
    booktitle   = {Proceedings of the Internet Measurement Conference},
    pages       = {36--49},
    year        = {2019}
}

@article{Claessen:2020:Reshaping,
    title       = {Reshaping the internet--the impact of the securitisation of internet infrastructure on approaches to internet governance: the case of Russia and the EU},
    author      = {Claessen, Eva},
    journal     = {Journal of Cyber Policy},
    volume      = {5},
    number      = {1},
    pages       = {140--157},
    year        = {2020},
    publisher   = {Taylor \& Francis}
}

@inproceedings{Kukkola:2018:Civilian,
    title           = {Civilian and military information infrastructure and the control of the Russian segment of Internet},
    author          = {Kukkola, Juha},
    booktitle       = {2018 International Conference on Military Communications and Information Systems (ICMCIS)},
    pages           = {1--8},
    year            = {2018},
    organization    = {IEEE}
}

@article{Ermoshina:2022:Market,
    title       = {A market of black boxes: The political economy of Internet surveillance and censorship in Russia},
    author      = {Ermoshina, Ksenia and Loveluck, Benjamin and Musiani, Francesca},
    journal     = {Journal of Information Technology \& Politics},
    volume      = {19},
    number      = {1},
    pages       = {18--33},
    year        = {2022},
    publisher   = {Taylor \& Francis}
}

@article{Vitak:2018:Knew,
    title       = {{`I Knew It Was Too Good to Be True'}: The Challenges Economically Disadvantaged Internet Users Face in Assessing Trustworthiness, Avoiding Scams, and Developing Self-Efficacy Online},
    author      = {Vitak, Jessica and Liao, Yuting and Subramaniam, Mega and Kumar, Priya},
    journal     = {Proceedings of the ACM on human-computer interaction},
    volume      = {2},
    number      = {CSCW},
    pages       = {1--25},
    year        = {2018},
    publisher   = {ACM New York, NY, USA}
}

@inproceedings{Bernstein:2013:Quantifying,
    title       = {Quantifying the invisible audience in social networks},
    author      = {Bernstein, Michael S and Bakshy, Eytan and Burke, Moira and Karrer, Brian},
    booktitle   = {Proceedings of the SIGCHI conference on human factors in computing systems},
    pages       = {21--30},
    year        = {2013}
}

@article{Christofides:2012:Hey,
    title       = {Hey mom, what’s on your Facebook? Comparing Facebook disclosure and privacy in adolescents and adults},
    author      = {Christofides, Emily and Muise, Amy and Desmarais, Serge},
    journal     = {Social Psychological and Personality Science},
    volume      = {3},
    number      = {1},
    pages       = {48--54},
    year        = {2012},
    publisher   = {Sage Publications Sage CA: Los Angeles, CA}
}

@article{Syn:2015:Social,
    title       = {Why do social network site users share information on Facebook and Twitter?},
    author      = {Syn, Sue Yeon and Oh, Sanghee},
    journal     = {Journal of Information Science},
    volume      = {41},
    number      = {5},
    pages       = {553--569},
    year        = {2015},
    publisher   = {SAGE Publications Sage UK: London, England}
}

@article{Lee:2015:Double,
    title       = {The double-edged sword: The effects of journalists' social media activities on audience perceptions of journalists and their news products},
    author      = {Lee, Jayeon},
    journal     = {Journal of Computer-Mediated Communication},
    volume      = {20},
    number      = {3},
    pages       = {312--329},
    year        = {2015},
    publisher   = {Oxford University Press Oxford, UK}
}

@article{Hagen:2018:Crisis,
    title       = {Crisis communications in the age of social media: A network analysis of Zika-related tweets},
    author      = {Hagen, Loni and Keller, Thomas and Neely, Stephen and DePaula, Nic and Robert-Cooperman, Claudia},
    journal     = {Social science computer review},
    volume      = {36},
    number      = {5},
    pages       = {523--541},
    year        = {2018},
    publisher   = {SAGE Publications Sage CA: Los Angeles, CA}
}

@inproceedings{Leavitt:2014:Upvoting,
    title       = {Upvoting hurricane Sandy: event-based news production processes on a social news site},
    author      = {Leavitt, Alex and Clark, Joshua A},
    booktitle   = {Proceedings of the SIGCHI conference on human factors in computing systems},
    pages       = {1495--1504},
    year        = {2014}
}

@article{Ozduzen:2020:Digital,
    title       = {Digital traces of “twitter revolutions”: Resistance, polarization, and surveillance via contested images and texts of occupy Gezi},
    author      = {Ozduzen, Ozge and McGarry, Aidan},
    journal     = {International Journal of Communication},
    volume      = {14},
    pages       = {2543--2563},
    year        = {2020},
    publisher   = {University of Southern California}
}

@article{Akbari:2019:Platform,
    title       = {Platform surveillance and resistance in Iran and Russia: The case of Telegram},
    author      = {Akbari, Azadeh and Gabdulhakov, Rashid},
    journal     = {Surveillance \& Society},
    volume      = {17},
    number      = {1/2},
    pages       = {223--231},
    year        = {2019}
}

@article{Gabdulhakov:2020:Trolling,
    title       = {(Con)trolling the Web: Social Media User Arrests, State-Supported Vigilantism and Citizen Counter-Forces in Russia},
    author      = {Gabdulhakov, Rashid},
    journal     = {Global Crime},
    volume      = {21},
    number      = {3-4},
    pages       = {283--305},
    year        = {2020},
    publisher   = {Taylor \& Francis}
}

@article{Sit:2019:Identifying,
    title       = {Identifying disaster-related tweets and their semantic, spatial and temporal context using deep learning, natural language processing and spatial analysis: a case study of Hurricane Irma},
    author      = {Sit, Muhammed Ali and Koylu, Caglar and Demir, Ibrahim},
    journal     = {International Journal of Digital Earth},
    year        = {2019},
    publisher   = {Taylor \& Francis}
}

@article{Depaula:2018:Information,
    title       = {Information strategies and affective reactions: How citizens interact with government social media content},
    author      = {DePaula, Nic and Dincelli, Ersin},
    journal     = {First Monday},
    year        = {2018}
}

@article{Imran:2015:Processing,
    title       = {Processing social media messages in mass emergency: A survey},
    author      = {Imran, Muhammad and Castillo, Carlos and Diaz, Fernando and Vieweg, Sarah},
    journal     = {ACM Computing Surveys (CSUR)},
    volume      = {47},
    number      = {4},
    pages       = {1--38},
    year        = {2015},
    publisher   = {ACM New York, NY, USA}
}

@inproceedings{Olteanu:2015:Expect,
    title       = {What to expect when the unexpected happens: Social media communications across crises},
    author      = {Olteanu, Alexandra and Vieweg, Sarah and Castillo, Carlos},
    booktitle   = {Proceedings of the 18th ACM conference on computer supported cooperative work \& social computing},
    pages       = {994--1009},
    year        = {2015}
}

@inproceedings{Xue:2021:Throttling,
    title       = {Throttling Twitter: an emerging censorship technique in Russia},
    author      = {Xue, Diwen and Ramesh, Reethika and Evdokimov, Leonid and Viktorov, Andrey and Jain, Arham and Wustrow, Eric and Basso, Simone and Ensafi, Roya},
    booktitle   = {Proceedings of the 21st ACM Internet Measurement Conference},
    pages       = {435--443},
    year        = {2021}
}

@article{Reuter:2018:Fifteen,
    title       = {Fifteen years of social media in emergencies: a retrospective review and future directions for crisis informatics},
    author      = {Reuter, Christian and Kaufhold, Marc-Andr{\'e}},
    journal     = {Journal of contingencies and crisis management},
    volume      = {26},
    number      = {1},
    pages       = {41--57},
    year        = {2018},
    publisher   = {Wiley Online Library}
}

@article{Reuter:2018:Social,
    title       = {Social media in crisis management: An evaluation and analysis of crisis informatics research},
    author      = {Reuter, Christian and Hughes, Amanda Lee and Kaufhold, Marc-Andr{\'e}},
    journal     = {International Journal of Human--Computer Interaction},
    volume      = {34},
    number      = {4},
    pages       = {280--294},
    year        = {2018},
    publisher   = {Taylor \& Francis}
}

@article{Wang:2018:Social,
    title       = {Social media analytics for natural disaster management},
    author      = {Wang, Zheye and Ye, Xinyue},
    journal     = {International Journal of Geographical Information Science},
    volume      = {32},
    number      = {1},
    pages       = {49--72},
    year        = {2018},
    publisher   = {Taylor \& Francis}
}

@inproceedings{Leavitt:2017:Role,
    title       = {The role of information visibility in network gatekeeping: Information aggregation on Reddit during crisis events},
    author      = {Leavitt, Alex and Robinson, John J},
    booktitle   = {Proceedings of the 2017 ACM conference on computer supported cooperative work and social computing},
    pages       = {1246--1261},
    year        = {2017}
}

@inproceedings{Temnikova:2015:Case,
    title       = {The case for readability of crisis communications in social media},
    author      = {Temnikova, Irina and Vieweg, Sarah and Castillo, Carlos},
    booktitle   = {Proceedings of the 24th international conference on world wide web},
    pages       = {1245--1250},
    year        = {2015}
}

@article{Zubiaga:2018:Detection,
    title       = {Detection and resolution of rumours in social media: A survey},
    author      = {Zubiaga, Arkaitz and Aker, Ahmet and Bontcheva, Kalina and Liakata, Maria and Procter, Rob},
    journal     = {ACM Computing Surveys (CSUR)},
    volume      = {51},
    number      = {2},
    pages       = {1--36},
    year        = {2018},
    publisher   = {ACM New York, NY, USA}
}

@inproceedings{Busse:2019:Snooze,
    author      = {Karoline Busse and Dominik Wermke and Sabrina Amft and Sascha Fahl and Emanuel von Zezschwitz and Matthew Smith},
    booktitle   = {Proceedings of the 2019 Workshop on Usable Security (USEC), USEC 2019, San Diego, CA, USA, February 24, 2019},
    doi         = {10.14722/usec.2019.23001},
    month       = feb,
    title       = {Replication: Do We Snooze If We Can't Lose? Modelling Risk with Incentives in Habituation User Studies},
    url         = {https://dx.doi.org/10.14722/usec.2019.23001},
    year        = {2019}
}

@inproceedings{alkhadra2021solarwinds,
  title={Solar winds hack: In-depth analysis and countermeasures},
  author={Alkhadra, Rahaf and Abuzaid, Joud and AlShammari, Mariam and Mohammad, Nazeeruddin},
  booktitle={2021 12th International Conference on Computing Communication and Networking Technologies (ICCCNT)},
  pages={1--7},
  year={2021},
  organization={IEEE}
}

@article{Corbin:1990:grounded,
	title     = {{Grounded theory research: Procedures, canons and evaluative criteria}},
	author    = {Corbin, Juliet and Strauss, Anselm},
	journal   = {Qualitative Sociology},
	volume    = {19},
	number    = {6},
	pages     = {418--427},
	year      = {1990}
}

@article{crosignani2021:SCattack,
  title={Pirates without borders: The propagation of cyberattacks through firms’ supply chains},
  author={Crosignani, Matteo and Macchiavelli, Marco and Silva, Andr{\'e} F},
  journal={FRB of New York Staff Report},
  number={937},
  year={2021}
}

@book{Strauss:1997:grounded,
	title     = {Grounded theory in practice},
	author    = {Strauss, Anselm and Corbin, Juliet M},
	pages     = {288},
	year      = {1997},
	publisher = {Sage}
}

@book{Charmaz:2014:grounded,
	title={{Constructing Grounded Theory}},
	author={Charmaz, Kathy},
	year={2014},
	publisher={Sage}
}

@misc{Acsc:2022:Advisory,
    title           = {Australian organisations encouraged to urgently adopt an enhanced cyber security posture},
    author          = {{Australian Cyber Security Center (ACSC)}},
    year            = {2022},
    month           = mar,
    day             = {28},
    howpublished    = {\url{https://www.cyber.gov.au/acsc/view-all-content/alerts/australian-organisations-encouraged-urgently-adopt-enhanced-cyber-security-posture}},
    note            = {Accessed 2022-06-02}
}

@misc{cisa:2022:shieldsup,
    title           = {{SHIELDS UP}},
    author          = {{Cybersecurity and Infrastructure Security Agency (CISA)}},
    year            = {2022},
    month           = may,
    day             = {10},
    howpublished    = {\url{https://www.cisa.gov/shields-up}},
    note            = {Accessed 2022-06-02}
}

@misc{ncsc:2022:encourage,
    title           = {UK organisations encouraged to take action in response to current situation in and around Ukraine},
    author          = {{National Cyber Security Centre (NCSC)}},
    year            = {2022},
    month           = jan,
    day             = {28},
    howpublished    = {\url{https://www.ncsc.gov.uk/news/uk-organisations-encouraged-to-take-action-around-ukraine-situation}},
    note            = {Accessed 2022-08-09}
}

@misc{bsi:2022:statement,
    title           = {{Einschätzung der aktuellen Cyber-Sicherheitslage in Deutschland nach dem russischen Angriff auf die Ukraine}},
    author          = {{Bundesamt für Informationssicherheit (BSI)}},
    year            = {2022},
    month           = may,
    day             = {12},
    howpublished    = {\url{https://www.bsi.bund.de/DE/Service-Navi/Presse/Pressemitteilungen/Presse2022/220225_Angriff-Ukraine-Statement.html?nn=1025778}},
    note            = {Accessed 2022-08-09}
}

@misc{can:2022:bulletin,
    title           = {Cyber threat bulletin: Cyber Centre reminds Canadian critical infrastructure operators to raise awareness and take mitigations against known Russian-backed cyber threat activity},
    author          = {{Canadian Centre for Cyber Security}},
    year            = {2022},
    month           = feb,
    day             = {13},
    howpublished    = {\url{https://cyber.gc.ca/en/guidance/cyber-threat-bulletin-cyber-centre-reminds-canadian-critical-infrastructure-operators}},
    note            = {Accessed 2022-08-09}
}

@misc{jakub2022:attacktimeline,
  title={Russia's war on Ukraine: Timeline of cyber-attacks},
  author={Jakub, PRZETACZNIK},
  year={2022},
  publisher={EPRS: European Parliamentary Research Service}
}

@misc{mehrotra-2022:signalfakenews,
    title           = {{Russia-Ukraine War: Signal Says Hack Claims Part Of 'coordinated Misinformation Campaign'}},
    author          = {{Shikhar Mehrotra)}},
    year            = {2022},
    month           = mar,
    day             = {1},
    howpublished    = {\url{https://www.republicworld.com/world-news/russia-ukraine-crisis/russia-ukraine-war-signal-says-hack-claims-part-of-coordinated-misinformation-campaign-articleshow.html}},
    note            = {Accessed 2022-07-27}
}

@misc{weinert-2019:mfaefficacy,
    title           = {{Your Pa\$\$word doesn't matter}},
    author          = {{Alex Weinert}},
    year            = {2019},
    month           = jul,
    day             = {9},
    howpublished    = {\url{https://techcommunity.microsoft.com/t5/microsoft-entra-azure-ad-blog/your-pa-word-doesn-t-matter/ba-p/731984}},
    note            = {Accessed 2022-08-18}
}

@article{Spearman:1987:correlation,
 ISSN = {00029556},
 URL = {http://www.jstor.org/stable/1422689},
 author = {C. Spearman},
 journal = {The American Journal of Psychology},
 number = {3/4},
 pages = {441--471},
 publisher = {University of Illinois Press},
 title = {The Proof and Measurement of Association between Two Things},
 urldate = {2022-08-02},
 volume = {100},
 year = {1987}
}

@article{Starbird:2019:Nature,
author = {Starbird, Kate},
year = {2019},
month = {07},
pages = {449-450},
title = {Disinformation’s spread: bots, trolls and all of us},
volume = {571},
journal = {Nature},
doi = {10.1038/d41586-019-02235-x}
}

@inproceedings{Starbird:2015:Connected,
author = {Huang, Y. Linlin and Starbird, Kate and Orand, Mania and Stanek, Stephanie A. and Pedersen, Heather T.},
title = {Connected Through Crisis: Emotional Proximity and the Spread of Misinformation Online},
year = {2015},
isbn = {9781450329224},
publisher = {Association for Computing Machinery},
address = {New York, NY, USA},
url = {https://doi.org/10.1145/2675133.2675202},
doi = {10.1145/2675133.2675202},
booktitle = {Proceedings of the 18th ACM Conference on Computer Supported Cooperative Work \& Social Computing},
pages = {969–980},
numpages = {12},
location = {Vancouver, BC, Canada},
series = {CSCW '15}
}

@article{Starbird:2018:BLM,
author = {Arif, Ahmer and Stewart, Leo Graiden and Starbird, Kate},
title = {Acting the Part: Examining Information Operations Within \#BlackLivesMatter Discourse},
year = {2018},
issue_date = {November 2018},
publisher = {Association for Computing Machinery},
address = {New York, NY, USA},
volume = {2},
number = {CSCW},
url = {https://doi.org/10.1145/3274289},
doi = {10.1145/3274289},
journal = {Proc. ACM Hum.-Comput. Interact.},
month = nov,
articleno = {20},
numpages = {27}
}

@article{Houlden:2021:Covid,
title = {The health belief model: How public health can address the misinformation crisis beyond COVID-19},
journal = {Public Health in Practice},
volume = {2},
pages = {100151},
year = {2021},
issn = {2666-5352},
doi = {https://doi.org/10.1016/j.puhip.2021.100151},
url = {https://www.sciencedirect.com/science/article/pii/S2666535221000768},
author = {Shandell Houlden and Jaigris Hodson and George Veletsianos and Darren Reid and Chris Thompson-Wagner},
}

@MISC{Schroeder:2012:Arab,
	note = {Combating Terrorism Exchange 2.4 (2012): 54-64},
	author = {Schroeder, Rob and Everton, Sean and Russell Shepherd},
	year = {2012},
	url = {http://hdl.handle.net/10945/53058},
	title = {Mining Twitter Data from the Arab Spring}
}

@INPROCEEDINGS{Bhalerao:2022:ethics,
  author={Bhalerao, Rasika and Hamilton, Vaughn and McDonald, Allison and Redmiles, Elissa M. and Strohmayer, Angelika},
  booktitle={2022 IEEE European Symposium on Security and Privacy Workshops (EuroS\&PW)}, 
  title={Ethical Practices for Security Research with At-Risk Populations}, 
  year={2022},
  volume={},
  number={},
  pages={546-553},
  doi={10.1109/EuroSPW55150.2022.00065}}

@article{McDonald:2019:IRR,
author = {McDonald, Nora and Schoenebeck, Sarita and Forte, Andrea},
title = {Reliability and Inter-Rater Reliability in Qualitative Research: Norms and Guidelines for CSCW and HCI Practice},
year = {2019},
issue_date = {November 2019},
publisher = {Association for Computing Machinery},
address = {New York, NY, USA},
volume = {3},
number = {CSCW},
url = {https://doi.org/10.1145/3359174},
doi = {10.1145/3359174},
journal = {Proc. ACM Hum.-Comput. Interact.},
month = nov,
articleno = {72},
numpages = {23}
}

@book{Beyer:1997:affinity,
author = {Beyer, Hugh and Holtzblatt, Karen},
title = {Contextual Design: Defining Customer-Centered Systems},
year = {1997},
isbn = {9780080503042},
publisher = {Morgan Kaufmann Publishers Inc.},
address = {San Francisco, CA, USA},
}

@inproceedings{Geeng:Fakenews:2020,
author = {Geeng, Christine and Yee, Savanna and Roesner, Franziska},
title = {Fake News on Facebook and Twitter: Investigating How People (Don't) Investigate},
year = {2020},
isbn = {9781450367080},
publisher = {Association for Computing Machinery},
address = {New York, NY, USA},
url = {https://doi.org/10.1145/3313831.3376784},
doi = {10.1145/3313831.3376784},
pages = {1–14},
numpages = {14},
location = {Honolulu, HI, USA},
series = {CHI '20}
}

@article{Starbird:Boston:2014,
author = {Starbird, Kate and Maddock, Jim and Orand, Mania and Achterman, Peg and Mason, Robert},
year = {2014},
month = {03},
pages = {},
title = {Rumors, False Flags, and Digital Vigilantes: Misinformation on Twitter After the 2013 Boston Marathon Bombing},
journal = {IConference},
doi = {10.9776/14308}
}
